\documentclass[12pt,a4paper]{article}

\usepackage{jheppub}
\graphicspath{{Img/}}

\usepackage{color}
\usepackage[dvipsnames, svgnames, x11names]{xcolor}
\usepackage[utf8]{inputenc}
\usepackage{float}
\usepackage{graphicx}
\usepackage{setspace}
\usepackage{subfigure}
\usepackage{bm}
\usepackage{bbm}
\usepackage{amsmath}
\usepackage{amssymb}
\usepackage{amsbsy}
\usepackage{amsthm}
\usepackage{amsfonts}
\usepackage{braket}
\usepackage{multirow}
\usepackage{ulem}
\usepackage{slashed}
\usepackage{dsfont}
\usepackage{csquotes}
\usepackage[compat=1.1.0]{tikz-feynman}

\usepackage{enumitem}
\newlist{senum}{enumerate}{1} 
\setlist[senum]{leftmargin=4.5cm ,align=left, label=\textbf{\Roman*}} 

\def\nn{\nonumber}

\def\be{\begin{equation}}       \def\ee{\end{equation}}
\def\bea{\begin{eqnarray}}      \def\eea{\end{eqnarray}}
\def\ba{\begin{array}}
	\def\ea{\end{array}}
\def\bnum{\begin{enumerate} }
	\def\enum{\end{enumerate}}

\def\nn{\nonumber}

\def\=>{\Rightarrow}
\def\>{\rightarrow}

\def\eye2{Fathbb{I}}

\title{Local Quenches from a Krylov Perspective}

\author{Pawel Caputa,$^{1,2,3}$}

\author{Giuseppe Di Giulio$^{1}$}

\preprint{YITP-25-26}

\affiliation[1]{
The Oscar Klein Centre and Department of Physics, Stockholm University, AlbaNova, 106 91 Stockholm, Sweden}
\affiliation[2]{Yukawa Institute for Theoretical Physics, Kyoto University, Kitashirakawa Oiwakecho, Sakyo-ku, Kyoto 606-8502, Japan}
\affiliation[3]{Faculty of Physics, University of Warsaw, Pasteura 5, 02-093 Warsaw, Poland}

\abstract{In this work, we investigate local quench dynamics in two-dimensional conformal field theories using Krylov space methods. We derive Lanczos coefficients, spread complexity, and Krylov entropies for local joining and splitting quenches in theories on an infinite line, a circle, a finite interval, and at finite temperature. We examine how these quantities depend on the central charge of the underlying conformal field theory and find that both spread complexity and Krylov entropy are proportional to it. Interestingly, Krylov entropies evolve logarithmically with time, mirroring standard entanglement entropies, making them useful for extracting the central charge.
In the large central charge limit, using holography, we establish a connection between the rate of spread complexity and the proper momentum of the tip of the end-of-the world brane, which probes the bulk analogously to a point particle.  Our results further demonstrate that spread complexity and Krylov entropies are powerful tools for probing non-equilibrium dynamics of interacting quantum systems.
}

\begin{document}
\sloppy
\maketitle

\newpage

\section{Introduction}

The dynamics of quantum systems is one of the most intriguing and actively studied areas of physics  \cite{eisert2015quantum,RevModPhys.83.863}. Surprisingly, many non-equilibrium phenomena such as thermalization \cite{Barthel2008,Cramer:2008zz,Cramer_2010,Calabrese_2012} or the lack thereof \cite{Nandkishore:2014kca,Abanin:2018yrt,Turner:2018kjz,Choi:2019wqq,Kormos:2016osj,Robinson:2018wbx}, dynamical phase transitions \cite{Heyl:2017blm}, and the interplay between quantum integrability and chaos \cite{Deutsch:1991msp,Srednicki:1994mfb,Rigol:2007juv,DAlessio:2015qtq}, remain poorly understood. Time-dependent process are also central to high-energy physics and quantum gravity. The key open problems include understanding thermalization in QCD \cite{Berges:2020fwq} and uncovering how and why black holes process and scramble quantum information at the fastest possible rate \cite{Sekino:2008he,Lashkari:2011yi}.  

A standard theoretical and experimental framework for studying non-equilibrium dynamics is the quench protocol \cite{Calabrese_2016}, where an initial state is prepared using a Hamiltonian $H_0$ and then evolved under a different Hamiltonian $H$. The Hamiltonians can differ from each other globally or locally and we refer to the corresponding scenarios as global or local quenches respectively. 
To probe these quench evolutions, quantum information tools such as Rényi and von Neumann entropies offer valuable insights. However, since entropies are typically associated with mixed rather than pure states, their definition requires specifying a subregion for which the reduced density matrix is computed.
Moreover, only a limited number of interesting cases can be solved analytically. Numerical methods and tensor network approaches, while useful, become less effective when a large amount of entanglement is generated during the evolution. 

Recently, a new set of tools and concepts has been developed to quantify the complexity of operators and states in many-body systems, quantum field theories (QFTs) and holography \cite{Roberts:2018mnp,Parker:2018yvk,Balasubramanian:2022tpr,Dymarsky:2021bjq}. The overarching goal for these efforts is to define a meaningful measure of the size of an operator under Heisenberg evolution or the spread of state under Schrodinger dynamics. The key tool in this framework is the Krylov basis (see Sec.\,\ref{sec:Preliminaries}), which allows us to expand an operator or state so that its dynamics can be effectively mapped onto a one-dimensional problem of a particle hopping on a chain. The average position of this particle along the chain is known as Krylov or spread complexity. Remarkably, this approach transforms quantum evolution into a probability distribution that fully encodes the system’s dynamics. This, in turn, allows for the definition of Krylov entropy, given by the Shannon entropy of this probability distribution, providing a natural notion of entropy for pure states without requiring the specification of sub-regions.

On the other hand, studies of Krylov and spread complexity have been largely limited to numerics, quantum mechanical models, or free QFTs, where the return amplitudes (crucial for computing Krylov complexity) are known. In this work, equipped with the Krylov toolbox, we revisit quantum quenches in 2D conformal field theories (CFTs) and apply these methods to gain new insights into quench dynamics as well as to Krylov data. In particular, we explore various local quench setups where return amplitudes, Lanczos coefficients, and the time evolution of spread complexity can be computed. Furthermore, we examine geometric setups where conformal maps enable analytic solutions. Notably, we derive the dependence of Lanczos coefficients and spread complexity on the central charge of the CFT.

Last but not least, within the AdS/CFT correspondence \cite{Maldacena:1997re}, measures of operator and state complexity in holographic CFTs are expected to provide new insights into gravity and black holes in Anti-de Sitter (AdS) spacetimes. Recently, the rate of spread complexity (its time derivative) in local operator quenches has been shown to correspond to the proper momentum of massive particles in $AdS_3$ spacetimes. In this work, we further explore this correspondence in more universal quench protocols, namely, the joining and splitting quenches, and analyze it within the AdS/BCFT construction \cite{Takayanagi:2011zk}.

This paper is organized as follows. In Sec.\,\ref{sec:Preliminaries}, we review the Krylov basis approach to complexity and quantum dynamics. In Sec.\,\ref{sec:LocalQ}, we discuss local quench protocols in 2D CFTs and compute return amplitudes in various examples. In Sec.\,\ref{sec:Krylovspacequantities}, we derive spread complexity and Krylov entropies for these local quenches and examine their dependence on the central charge. In Sec.\,\ref{sec:Holography}, we explore the holographic correspondence between the spread complexity rate and proper momentum, using massive particles and the AdS/BCFT setups. Finally we conclude in Sec.\,\ref{sec:Conclusions} and provide additional technical details in three appendices.
\section{Preliminaries}\label{sec:Preliminaries}
We start by reviewing basics of the Krylov basis approach to complexity. We provide a slightly more pedagogical summary for the quantum many-body oriented readers. More pedagogical treatment can be found in \cite{LanczosBook} and readers familiar with these tools can immediately proceed to the following section.
\subsection{Spread complexity and K-entropy}
Consider a quantum system prepared in an initial state $|\psi(0)\rangle$. For $t>0$, we let the system evolve through a time-independent Hamiltonian $H$ as
\begin{equation}
\label{eq:original_evolution}
    |\psi(t)\rangle=e^{- {\rm i} H t}|\psi(0)\rangle\,.
\end{equation}
Intuitively, the Krylov-basis method to quantify the complexity of the state at time $t$ consists of keeping track of the evolution in a given basis and counting how many basis vectors are visited at time $t$ and with what probability.
For this purpose, we consider a generic orthonormal basis 
$\mathcal{B}=\{\vert B_n\rangle,\,n=0,1,2,\dots \vert\mathcal{B} \vert \,\}$
and we denote by $\vert\mathcal{B} \vert$ the number of its elements.
The state \eqref{eq:original_evolution} can be expanded over the basis $\mathcal{B}$ as
\begin{equation}
|\psi(t)\rangle=\sum_{n}\langle B_n\vert \psi(t)\rangle\, \vert B_n\rangle\,.
\end{equation}
The quantity  
\begin{equation}
\label{eq:costfunction}
C_\mathcal{B}(t)=\sum_{n}n\vert\langle B_n\vert \psi(t)\rangle\vert^2\,,
\end{equation}
is the average value of the label $n$ based on the probability that the state $|\psi(t)\rangle$ is in the state $\vert B_n\rangle$ at time $t$. If the label $n$ in the basis $\mathcal{B}$ measures how much $\vert B_n\rangle$ is, in some sense, hard to obtain, \eqref{eq:costfunction} would be a meaningful notion of complexity. 

Since its introduction in computer science, complexity is tied to an optimization procedure. Applied to the case considered here, this suggests to define the {\it spread complexity} of the state $|\psi(t)\rangle$ as \cite{Balasubramanian:2022tpr}
\begin{equation}
\label{eq:spreadcompl_min}
C(t)=\min_\mathcal{B}\left(\sum_{n}n\vert\langle B_n\vert \psi(t)\rangle\vert^2\right)\,,
\end{equation}
where the minimization is performed over all the choices of basis.
In \cite{Balasubramanian:2022tpr}, it has been shown that the minimum in
\eqref{eq:spreadcompl_min}
is attained when $\mathcal{B}$ is the so-called Krylov basis.
The Krylov basis, denoted as
$\mathcal{K}=\{\vert K_n\rangle,\,n=0,1,2,\dots,\vert\mathcal{K} \vert\}$, is obtained through a Gram-Schmidt orthonormalization on the set of vectors $\{\vert \psi(0)\rangle,H\vert \psi(0)\rangle,H^2\vert \psi(0)\rangle,\dots,H^n\vert \psi(0)\rangle,\dots\}$. When applied in this context, this procedure is called the Lanczos algorithm \cite{Lanczos:1950zz}. The number $\vert\mathcal{K} \vert$ of vectors in the Krylov basis is not known a priori and has to satisfy $\vert\mathcal{K} \vert\leq \dim \mathcal{H}$.
Once determined the Krylov basis, the state \eqref{eq:original_evolution} can be expanded as
\begin{equation}
\label{eq:Krylov expansion}
|\psi(t)\rangle=\sum_{n}\psi_n(t)\, \vert K_n\rangle\,,
\qquad
\psi_n(t)\equiv\langle K_n\vert \psi(t)\rangle\,,
\end{equation}
and its spread complexity is given by
\begin{equation}
\label{eq:spread complexity}
C_\mathcal{K}(t)=\sum_{n} n\, p_n(t)\,,
\qquad
p_n(t)\equiv\vert\psi_n(t)\vert^2\,.
\end{equation}
This quantity is a generalization of the Krylov complexity for operators \cite{Parker:2018yvk} applied to time-evolving states.

 By construction, the vector $\vert K_n\rangle$ contains up to the $n$-th power of the Hamiltonian. Thus, the larger $n$, the higher the powers of $H$ involved and, therefore, the more complex $\vert K_n\rangle$. This characterizes the average value of $n$ in \eqref{eq:spread complexity} as a meaningful and useful complexity measure. Indeed, the Krylov/spread complexity has been successfully employed in several contexts, including quantum chaos \cite{Parker:2018yvk,Rabinovici:2022beu,Balasubramanian:2022dnj,He:2022ryk,Erdmenger:2023wjg}, topological phases of matter \cite{Caputa:2022eye,Caputa:2022yju}, quantum field theories \cite{Avdoshkin:2022xuw,Camargo:2022rnt,Caputa:2021ori,Malvimat:2024vhr}, open quantum systems and non-unitary dynamics \cite{Liu:2022god,Bhattacharya:2022gbz} as well as some types of time-dependent scenarios \cite{Takahashi:2024hex}. See review \cite{Nandy:2024htc} for more developments and other relevant references. In holography, the Krylov basis and complexity became instrumental in double-scaled SYK and Jackiw–Teitelboim gravity developments \cite{Lin:2022rbf,Berkooz:2018jqr,Berkooz:2018qkz} and in comparing complexity with volume of the Einstein-Rosen bridge \cite{Rabinovici:2023yex,Balasubramanian:2024lqk,Heller:2024ldz,Miyaji:2025yvm}. Moreover, recently, spread complexity has also provided a quantitative match with radial momentum when probing the evolution after an operator quench \cite{Caputa:2024sux}. 

The function $p_n(t)$ in \eqref{eq:spread complexity} is the probability that $|\psi(t)\rangle$ is in the state $\vert K_n\rangle$ at time $t$ and is the key object to determine in the Krylov-space approach to complexity.
Its knowledge for any $n$ allows us to access a further quantity, the Krylov entropy (K-entropy for short) \cite{Barbon:2019wsy}, which probes the spread of the state \eqref{eq:original_evolution} (see also \cite{Patramanis:2021lkx}). It is given by the Shannon entropy, i.e. the information content, of the probability distribution $p_n(t)$ and reads
\begin{equation}
\label{eq:Kentropy_def}
H_\mathcal{K}(t)=-\sum_{n}  p_n(t)\ln p_n(t)\,.
\end{equation}
Notice that, differently from the entanglement entropy, \eqref{eq:Kentropy_def} is dependent on the choice of the basis of $|\psi(t)\rangle$ and does not require any bipartition of the quantum system. Thus, on general grounds, these two quantities are not expected to have the same behaviour and can probe the dynamics complementarily.
\subsection{Lanczos coefficients and return amplitude}
Now, we summarize an efficient way to determine the probabilities $p_n(t)$ that we are interested in for computing Krylov-space quantities (see \cite{LanczosBook} for details).
The starting point is the fact that the Lanczos algorithm \cite{Lanczos:1950zz} can be defined by the tri-diagonal action of the Hamiltonian on the vectors of Krylov basis $\mathcal{K}$ 
\begin{equation}
\label{eq:tridiagonalaction}
  H\vert K_n\rangle= a_n \vert K_n\rangle+b_{n}\vert K_{n-1}\rangle+b_{n+1}\vert K_{n+1}\rangle\,.
\end{equation}
The coefficients in \eqref{eq:tridiagonalaction} are derived iteratively, and are known as Lanczos coefficients. Using the Schrödinger equation with \eqref{eq:tridiagonalaction} and the expansion in \eqref{eq:Krylov expansion}, we derive the evolution equation for the amplitudes $\psi_n(t)$
\begin{equation}
\label{eq:Schroedinger eq_amplitude}
    {\rm i} \partial_t \psi_n(t)= a_n \psi_n(t)+b_{n+1}\psi_{n+1}(t)+b_{n}\psi_{n-1}(t)\,.
\end{equation}
This procedure teaches us an insightful lesson; any quantum dynamics can be recast into effective one-dimensional dynamics in the Krylov space generated by a tridiagonal Hamiltonian. If the Lanczos coefficients are known, we can solve (at least numerically) \eqref{eq:Schroedinger eq_amplitude}, obtain the amplitudes, and finally the probabilities $p_n(t)$.

Performing explicitly the orthonormalization to obtain Krylov basis and, from this, the Lanczos coefficients is often hard, particularly in Hilbert spaces with large dimensionalities. For this reason, it is helpful to have a procedure bypassing the direct computation of the $\vert K_n\rangle$ and obtain
$a_n$ and $b_n$. Fortunately, Lanczos coefficients can be extracted directly from the return amplitude (sometimes called survival amplitude or Loschmidt amplitude) of the evolution considered
\begin{equation}
\label{eq:returnamplitude_def}
    S(t)\equiv \langle \psi(t)\vert \psi(0)\rangle=\langle \psi(0)\vert e^{ {\rm i} H t}\vert \psi(0)\rangle=\psi_0^*(t)\,.
\end{equation}
More specifically, the Lanczos coefficients are related to the moments of the return amplitude \eqref{eq:returnamplitude_def}
\begin{equation}
  \mu_n \equiv \frac{d^n}{dt^n} S(t)\Bigg\vert_{t=0}=\langle K_0\vert({\rm i}H)^n \vert K_0\rangle\,,
\end{equation}
through an easily implementable iterative procedure. We refer the reader to \cite{Balasubramanian:2022tpr} for details.

In summary, an efficient approach to obtaining the spread complexity involves computing the return amplitude \eqref{eq:returnamplitude_def} as a function of time, extracting the Lanczos coefficients from its moments, and solving \eqref{eq:Schroedinger eq_amplitude} to determine the amplitudes $\psi_n(t)$. These amplitudes are then used to compute the corresponding probabilities, which are substituted into \eqref{eq:spread complexity} to evaluate the spread complexity.
In the following sections, we will compute the return amplitudes for various types of local quenches and analyze the resulting evolution of the spread complexity and the K-entropies.

\section{Local quenches in CFT}\label{sec:LocalQ}
In the local quench protocol, we prepare a one-dimensional quantum system in an initial state $|\psi(0)\rangle$, and we let it evolve for $t>0$ through a Hamiltonian $H$ that modifies $|\psi(0)\rangle$ only at a few points. In this section, we review the CFT description of different types of local quench dynamics, compute the return amplitudes in these cases, and comment on existing results on the evolution of the entanglement entropies of the half system.
\subsection{Local quench protocols}
\label{subsec:protocols}
\begin{figure}[t!]
\centering
\includegraphics[width=1.\textwidth]{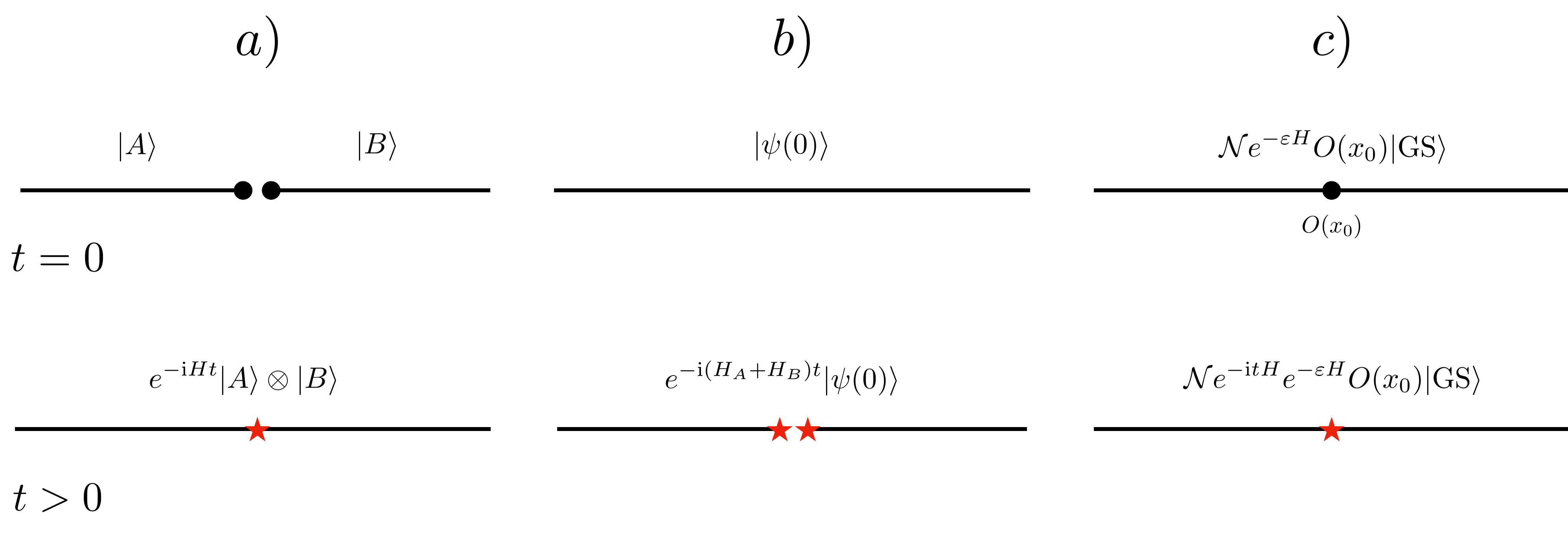}
\caption{Pictorial representation of the three local quenches discussed in this work. a) {\it Joining quench}: The initial state is the product of two states $|A\rangle$ and $|B\rangle$ on two complementary intervals, and we let the system evolve via a Hamiltonian, which couples them.
b) {\it Splitting quench}: The initial state is homogeneous, and the evolution Hamiltonian $H_A+H_B$ splits the system into two complementary parts.
c) {\it Operator quench}: The initial state is the ground state of a Hamiltonian $H$ locally excited by an operator $O(x_0)$, and the system is evolved through $H$.
}
\label{fig:protocols}
\end{figure}

We first provide an overview of the types of local quenches explored in this manuscript. In the considered cases, the evolution Hamiltonian changes the initial state in a single point. A pictorial representation of these quench protocols is reported in Fig.\,\ref{fig:protocols}.

\subsubsection{Joining quench}
\label{subsec:joiningquench_def}
Consider a one-dimensional system made by two connected regions $A$ and $B$, initially decoupled. Thus, we write the initial state as
\begin{equation}
|\psi(0)\rangle=|A\rangle\otimes |B\rangle\,,
\end{equation}
and we let it evolve as in \eqref{eq:original_evolution} through a Hamiltonian that couples $A$ and $B$. We can represent the evolution Hamiltonian as $H= H_A+ H_B+H_{AB}$, where $H_A$ and $H_B$ act non-trivially only on $A$ and $B$ respectively, while $H_{AB}$ couples the two parties.
Due to the coupling between initially decoupled regions, the resulting dynamics is non-trivial. We refer to this protocol as the {\it joining quench}, and we report a sketch of it in Fig.\,\ref{fig:protocols}a).
Historically, the joining quench was the first type of local quench considered in the literature \cite{Eisler_2007,Calabrese:2007mtj}. Since then, the post-quench evolution of several observables, including correlation functions and entanglement measures, has been studied in quantum chains \cite{Eisler_2008,Igloi_2009,Divakaran:2011bw,Eisler:2014dze,Alba_2014,DiGiulio:2021noo}, quantum field theories \cite{Cardy:2011zz,Stephan:2011kcw,Asplund:2013zba,Wen:2015qwa,Cardy:2016fqc}, and holographic scenarios \cite{Nozaki:2013wia,Ugajin:2013xxa,Astaneh:2014fga}.

In this manuscript, we focus on instances where the initially decoupled parties are states of critical systems and have a continuum description via CFT states. The evolution Hamiltonian is also assumed to be critical, becoming in the continuum limit a CFT Hamiltonian. Under these assumptions, the dynamics can be studied with CFT tools, as first pointed out in \cite{Calabrese:2007mtj}.
\subsubsection{Splitting quench}
In the joining quench protocol, two initially separated subsystems are brought together at time $t=0$. Conversely, we can consider a protocol that represents the opposite process. We assume the initial state is homogeneous and, at time $t=0$, we apply the evolution \eqref{eq:original_evolution} with Hamiltonian
\begin{equation}
\label{eq:splittingevolution}
H = H_A+H_B\,,
\end{equation}
where $H_A$ acts non-trivially on a connected region $A$ and $H_B$ on the complementary region $B$, also assumed to be connected. 
This protocol amounts to breaking the system into two complementary parties and letting them evolve independently. For this reason, this protocol is known as {\it splitting quench protocol} \cite{Shimaji:2018czt}. A pictorial representation of the splitting quench is shown in Fig.\,\ref{fig:protocols}b). This quench was introduced in \cite{Shimaji:2018czt}, where a CFT description and its holographic dual were discussed, together with the evolution of the entanglement entropy (see also \cite{Caputa:2019avh,Lap:2024hsy,Lap:2024vwm} for developments in this context).

In this formulation of the splitting quench, we create a single breaking point in the system at $t=0$, which separates the intervals $A$ and $B$. This protocol can be generalized to a multiple splitting quench, where the system is decoupled into $n\geq 2$ intervals at $t=0$. In this manuscript, we investigate only the single splitting quench and its comparison with the joining quench protocol described in Sec.\,\ref{subsec:joiningquench_def}.
In the CFT formulation we study in this paper, the initial state is a CFT state, while $H_A$ and $H_B$ in the evolution Hamiltonian \eqref{eq:splittingevolution}
are CFT Hamiltonians defined on the two intervals $A$ and $B$ separately.
\subsubsection{Operator quench}
\label{subsec:operatorquench}
Finally, we can consider a milder version of the local quench protocols above. Given the ground state $\vert\rm GS\rangle$ of a Hamiltonian $H$, the initial state is assumed to be the locally excited state
\begin{equation}
\label{eq:initialstate_localopquench}
    \vert\psi(0)\rangle=\mathcal N e^{-\varepsilon H}O(x_0)\vert\rm GS\rangle\,,
\end{equation}
where $O(x_0)$ is a local operator at a spatial point $x_0$ and $\mathcal N$ is a normalization constant. Since we consider a continuum system described by a quantum field theory, a regulator (smearing) $\varepsilon$ is introduced to ensure that the state $\vert\psi(0)\rangle$ has finite energy. We let the system evolve through the Hamiltonian $H$. Since $H$ and $O(x_0)$ do not commute in general, the induced evolution 
\begin{equation}
    \vert\psi(t)\rangle=\mathcal{N}e^{-{\rm i}t H}e^{-\varepsilon H}O(x_0)\vert {\rm GS}\rangle
    =\mathcal{N} O(x_0,t+{\rm i\varepsilon})\vert\rm GS\rangle\,,
\end{equation}
is non-trivial. We refer to this protocol as {\it local operator quench} and its graphical representation is given in Fig.\,\ref{fig:protocols}c).
In this manuscript, we focus on the local operator quench dynamics in 2D CFT, namely when $\vert\rm GS\rangle$ is a CFT ground state and $H$ a CFT Hamiltonian. We also assume that $O(x_0)$ is a primary operator with scaling dimension $\Delta=h+\bar{h}$. This CFT protocol has been introduced in \cite{Nozaki:2014hna}, where a holographic prescription for its dynamics was also given. The evolution of various observables, including entanglement measures, was studied in several subsequent works \cite{Caputa:2014vaa, Nozaki:2014uaa,Asplund:2014coa,Caputa:2015qbk,Jahn:2017xsg,Kudler-Flam:2020yml,Kudler-Flam:2020xqu}. In the coming sections, we also explore the instance where the initial state is a CFT thermal state locally excited by a primary operator.
\subsection{Return amplitudes after local quenches in CFT}\label{subsec:returnampl_CFT}
Next, we discuss the time evolution of return amplitudes along the local quench dynamics in 2D CFTs. We will review some known results and derive new ones focusing on the types of local quenches discussed in Sec.\,\ref{subsec:protocols}.
\subsubsection{Joining quench}
\label{subsec:S_joiningquench}

Consider the setup described in Sec.\,\ref{subsec:joiningquench_def}. We assume that the initially separated parties are in the ground states of critical Hamiltonians, which, in the continuum limit, are described by the same CFT with central charge $c$. Our focus is on CFTs in Minkowski spacetime. However, during the computations, we work mainly in Euclidean time $\tau=-{\rm i}t$, and we go back to Lorentzian time through Wick's rotation only once the expression of the return amplitude is obtained.
The scenario we study here consists of two initially decoupled ground states of two identical CFTs on an interval of equal length $L/2$ with Hamiltonian $H_{L/2}$. Since the CFTs are defined on spaces with boundaries, it is more appropriate to consider them as boundary CFTs (BCFTs). The presence of boundaries is crucial for the joining quench dynamics. At the same time, the CFT's boundary conditions are irrelevant since distinct boundary conditions give different non-universal contributions.
The evolution after the joining quench occurs through the CFT Hamiltonian $H_{L}$, which is the same inducing the two initial ground states but defined on the union of the two intervals. Thus, the total length of the evolving systems is $L$.
Following the notations of Sec.\,\ref{subsec:joiningquench_def}, we have that
\begin{equation}
  H_A=H_{L/2}\otimes\boldsymbol{1}_B  \,,
  \qquad
  H_B=\boldsymbol{1}_A\otimes H_{L/2}  \,,
  \qquad
  H_A+H_B+H_{AB}=H_{L}\,,
\end{equation}
where $\boldsymbol{1}_A$ and $\boldsymbol{1}_B$ act as the identity on the intervals $A$ and $B$ respectively.
We review here the computation of the return amplitude after this joining quench as reported in \cite{Stephan:2011kcw}. Despite this result being known, detailing here the calculation is helpful to extend the results to other local quenches, as we discuss in the coming subsections.

\begin{figure}[t!]
\centering
\includegraphics[width=1.\textwidth]{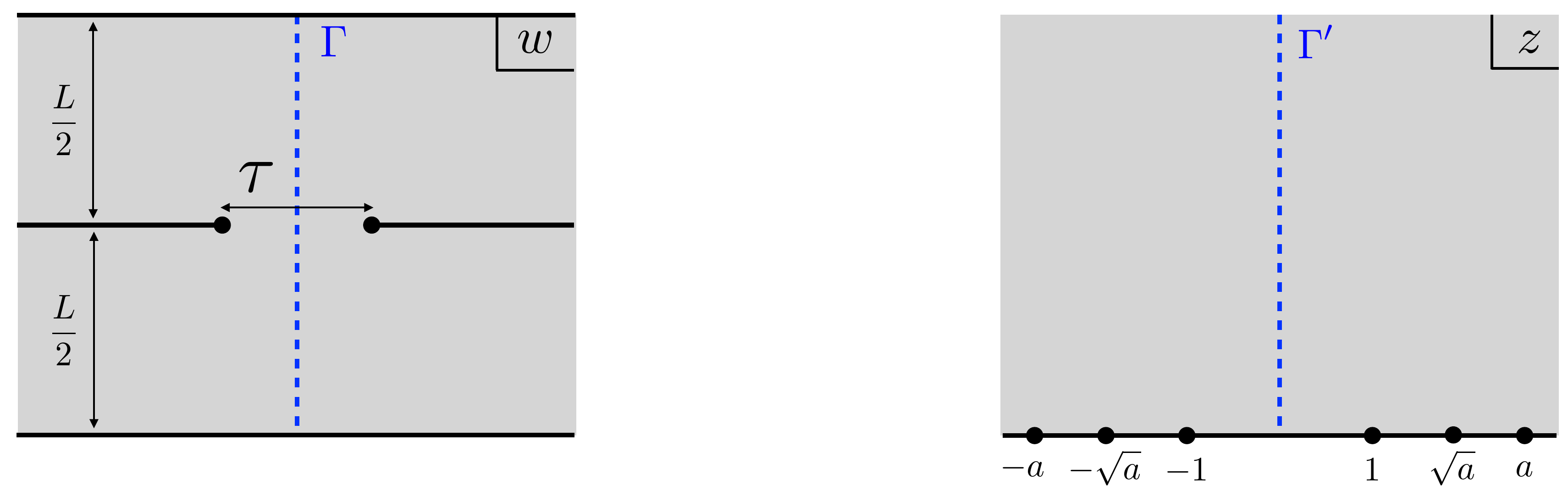}
\caption{In the left panel, we show the two-dimensional geometry whose CFT partition function gives the return amplitude after a joining quench of two intervals of size $L/2$. This geometry can be mapped to the half-plane shown in the right panel. The expression of $a$ in terms of $\tau$ and $L$ is given in \eqref{eq:def_parameter_a}. The curves $\Gamma$ and $\Gamma'$ used to compute the free energy \eqref{eq:variationF_Schwarz} are represented as blue dashed lines.}
\label{fig:JoiningQuenchGeom}
\end{figure}

Given that the states $|A\rangle$ and $|B\rangle$ are both ground states of $H_{L/2}$, the initial state can be obtained by projecting a generic state $|s\rangle$ defined on the full interval $A\cup B$ as follows \cite{Stephan:2011kcw}
\begin{equation}
\label{eq:initialstate_projection}
  |A\rangle\otimes |B\rangle\propto \lim_{\lambda\to\infty}e^{-\lambda(H_{L/2}\otimes \boldsymbol{1}_B+\boldsymbol{1}_A\otimes H_{L/2})}|s\rangle  \,.
\end{equation}
The choice of the state $|s\rangle$ is arbitrary (provided that $\langle s|e^{-\lambda(H_{L/2}\otimes \boldsymbol{1}_B+\boldsymbol{1}_A\otimes H_{L/2})}|s\rangle\neq 0$), given that different states would give different normalization factors.
The expression \eqref{eq:initialstate_projection} allows to write the (Euclidean) return amplitude \eqref{eq:returnamplitude_def} as
\begin{equation}
\label{eq:returnamplitudeEuclidean}
   S(\tau)\propto  \lim_{\lambda\to\infty}\langle s|e^{-\lambda(H_{L/2}\otimes \boldsymbol{1}_B+\boldsymbol{1}_A\otimes H_{L/2})}e^{-\tau H}e^{-\lambda(H_{L/2}\otimes \boldsymbol{1}_B+\boldsymbol{1}_A\otimes H_{L/2})}|s\rangle\equiv Z(\tau)\,.
\end{equation}
The normalization factor omitted in (\ref{eq:returnamplitudeEuclidean}) ensures that, after Wick's rotation, $S(t=0)=1$.
In the CFT language, $Z(\tau)$ in \eqref{eq:returnamplitudeEuclidean}
is interpreted as the partition function on the geometry represented in the left panel of Fig.\,\ref{fig:JoiningQuenchGeom} \cite{Stephan:2011kcw}. In particular, the semi-infinite left and the right parts (with three horizontal thick black lines) are generated by $e^{-\lambda(H_{L/2}\otimes \boldsymbol{1}_B+\boldsymbol{1}_A\otimes H_{L/2})}$, while the central one (with only two black lines) is generated by $e^{-\tau H}$.
The strategy we follow for determining the partition function $Z(\tau)$ is to compute the corresponding CFT free energy
\begin{equation}
\label{eq:freeenergydef}
\mathcal{F}(\tau)=-\ln Z(\tau)\,.
\end{equation}
This can be obtained by evaluating its infinitesimal variation under a conformal transformation and exploiting the mapping between the CFT geometry of the problem and the half-plane \cite{Cardy:1988tk,DiFrancesco:1997nk}. As we will see in the coming sections, we can apply this method to all those local quench geometries that can be mapped to the half-plane.

Let us call $w=x_0+{\rm i} x_1$ the complex coordinate of the original local quench geometry in the left panel of Fig.\,\ref{fig:JoiningQuenchGeom}. Let us denote by $\Gamma$ the vertical line located along the imaginary axis. We perform an infinitesimal conformal transformation defined as follows
\begin{equation}
w
\to
\begin{cases}
 w+\delta\tau   & {\rm Re}w>0   \,, \\
 w   & {\rm Re}w<0\,,
\end{cases}
\end{equation}
which induces an infinitesimal variation of the free energy which reads \cite{DiFrancesco:1997nk, Stephan:2011kcw}
\begin{equation}
\label{eq:variationFreeenergy}
\delta\mathcal{F}=\frac{\delta\tau}{2\pi{\rm i}}\int_\Gamma d w \langle T(w)\rangle + {\rm c.c.} \,,
\end{equation}
where the complex conjugate comes from the contribution of the anti-holomorphic part of the stress-energy tensor.
The geometry in the left panel of Fig.\,\ref{fig:JoiningQuenchGeom} can be mapped to the half-plane with complex coordinate $z$ (right panel of Fig.\,\ref{fig:JoiningQuenchGeom}) via the transformation
\begin{equation}
\label{eq:from UHPtoTrouser}
    w(z)=\frac{L}{2\pi}\left[\ln\left(\frac{z-1}{z+1}\right)+\ln\left(\frac{z-a}{z+a}\right)-{\rm i} \pi\right]\,,
\end{equation}
where
\begin{equation}
\label{eq:def_parameter_a}
   a=\left[\coth\frac{\pi \tau}{4L}\right]^2 \,.
\end{equation}
Using the transformation rule of the expectation value of the stress-energy tensor and the fact that it is zero on the upper-half plane, \eqref{eq:variationFreeenergy} leads to
\begin{equation}
\label{eq:variationF_Schwarz}
\delta\mathcal{F}=
-\frac{c\delta\tau}{24\pi {\rm i}}\int_{\Gamma'} d z \sigma(z) + {\rm c.c.} \,,
\end{equation}
where $\sigma$ is the following composition of maps
\begin{equation}
\label{eq:def_sigma} 
    \sigma(z)\equiv \{w(z),z\}(w'(z))^{-1}\,,
\end{equation} 
and the Schwarzian derivative is defined as
\begin{equation}
\label{eq:defSchwarzian}
\{w(z),z\}=\frac{w'''}{w'}-\frac{3}{2}\left(\frac{w''}{w'}\right)^2\,.
\end{equation}
The integration domain $\Gamma'$ in \eqref{eq:variationF_Schwarz} is the curve on the half-plane where
the axis $\Gamma$ is mapped.
Using \eqref{eq:from UHPtoTrouser} in \eqref{eq:def_sigma}, we observe that $\sigma(z)$ has poles in $z=\pm 1$, $z=\pm \sqrt{a}$, and $z=\pm a$.
Since $\Gamma$ is mapped into $\Gamma'$ along the imaginary axis, the integral can be computed by extending it to the full vertical axis and then closing the contour at infinity in the right part of the plane. By the residue theorem, we have
\begin{eqnarray}
\nonumber
\label{eq:variationF_joining_v2}
\delta\mathcal{F}&=&
\frac{c\delta\tau}{24}\left(\mathrm{Res}\left[\sigma(z);z=1\right]+\mathrm{Res}\left[\sigma(z);z=a\right]+\mathrm{Res}\left[\sigma(z);z=\sqrt{a}\right]\right)+ {\rm c.c.}
\\
&=&
-\frac{\pi c\delta\tau}{6L}+\frac{\pi c\delta\tau}{32L}\frac{a^2+6a+1}{(1+a)\sqrt{a}}
\,.
\end{eqnarray}
Integrating in $\tau$ and taking the exponential, we find
\begin{equation}
\label{eq:Z_joining_v1}
Z(\tau)\propto e^{\frac{\pi c\tau}{6L}}\left[\sinh\left(\frac{\pi \tau}{L}\right)\right]^{-c/8}\,.
\end{equation}
According to \eqref{eq:returnamplitudeEuclidean}, the Euclidean return amplitude is proportional to the right-hand side of \eqref{eq:Z_joining_v1}. Performing the Wick's rotation
$\tau=-{\rm i}t+2\varepsilon$, we can write \cite{Stephan:2011kcw}
\begin{equation}
\label{eq:returnamplitude_realtime}
  S(t)\propto e^{-{\rm i}\frac{\pi c t}{6 L}}\left[-{\rm i}\sin\left(\frac{\pi (t+2{\rm i}\varepsilon)}{L}\right)\right]^{-c/8}   \,.
\end{equation}
Here, the parameter $\varepsilon$ plays the role of an ultraviolet (UV) cutoff, needed to avoid divergences when $t=0$. Thus, the CFT predictions for the return amplitude are reliable as long as $t\gg2\varepsilon $. Normalizing the expression in \eqref{eq:returnamplitude_realtime} to ensure that $S(t=0)=1$, we obtain
\begin{equation}
\label{eq:Snormalized}
  S(t)=
e^{-{\rm i}\frac{\pi c t}{6 L}}\left[\frac{\sin\left(\frac{\pi (t+2{\rm i}\varepsilon)}{L}\right)}{\sin\left(\frac{2\pi {\rm i}\varepsilon}{L}\right)}\right]^{-c/8}.
\end{equation}
An alternative derivation of this result is reported in Appendix \ref{app:CFTdetails}, where different CFT techniques are considered and exploited to extend \eqref{eq:Snormalized} to other interesting cases.
In Sec.\,\ref{sec:Krylovspacequantities}, we will exploit the expression \eqref{eq:Snormalized} to determine the spread complexity and the K-entropy evolutions after a local joining quench. 
Notice that the multiplicative phase in \eqref{eq:Snormalized} can be, in principle, removed by shifting the ground state energy in the system, and, therefore, we do not expect to play any role in physically relevant quantities. We will confirm this expectation when we investigate the Krylov-space dynamics of this process in the coming sections. Following this remark, from now on, we will get rid of this phase factor with the appropriate ground state energy shift unless otherwise specified. 

We find it worth commenting on the limit $L\to \infty$ of the analytical results reported above. This amounts to considering a joining quench where the initially decoupled CFT ground states are defined on two semi-infinite lines. The variation \eqref{eq:variationF_joining_v2} of the free energy becomes in this limit
\begin{equation}
\label{eq:variationF_joining_infinite}
\delta\mathcal{F}=\frac{ c\delta\tau}{8\tau}\,.
\end{equation}
Integrating and taking the exponential it is straightforward to obtain
\cite{Stephan:2011kcw}
\begin{equation}
\label{eq:Paritition_joining_infinite}
Z(\tau)\propto  \tau^{-c/8}\,,
\end{equation}
which, using \eqref{eq:returnamplitudeEuclidean}, leads to the properly normalized return amplitude
\begin{equation}
\label{eq:Snormalized_infinite}
  S(t)= 
  \left(\frac{t+2{\rm i}\varepsilon}{2{\rm i} \varepsilon}\right)^{-c/8}
  \,.
\end{equation}
As pointed out above in the finite-size case, also in this limit, the CFT predictions are reliable for times much larger than the cutoff $\varepsilon$.

We conclude this subsection by discussing another scenario for which the return amplitude can be deduced from \eqref{eq:Snormalized}. At time $t=0$, we consider two identical thermal CFTs at the same temperature $\beta^{-1}$ defined on semi-infinite lines. At time $t>0$, we let the system evolve through a CFT Hamiltonian, which couples the two semi-infinite lines. This is a local joining quench of a finite-temperature CFT. To derive the return amplitude, we can use the fact that finite-temperature CFTs on infinite lines are described on a cylindrical geometry, where the compact direction is the Euclidean time and has size $\beta$. Thus, we can exploit the return amplitude \eqref{eq:Snormalized} after a local joining quench of a finite-size CFT and substitute $L\to {\rm i}\beta$. The resulting return amplitude reads
 \begin{equation}
    \label{eq:SfiniteT}
        S(t)=
  \left[\frac{\sinh\left(\frac{\pi (t+2{\rm i}\varepsilon)}{\beta}\right)}{\sinh\left(\frac{2\pi {\rm i}\varepsilon}{\beta}\right)}\right]^{-c/8}
  \,.
    \end{equation}
    In Appendix \ref{subapp:otherquenches}, we give the details of the derivation of \eqref{eq:SfiniteT}.

\begin{figure}[t!]
\hspace{0.15cm}
\includegraphics[width=.45\textwidth]{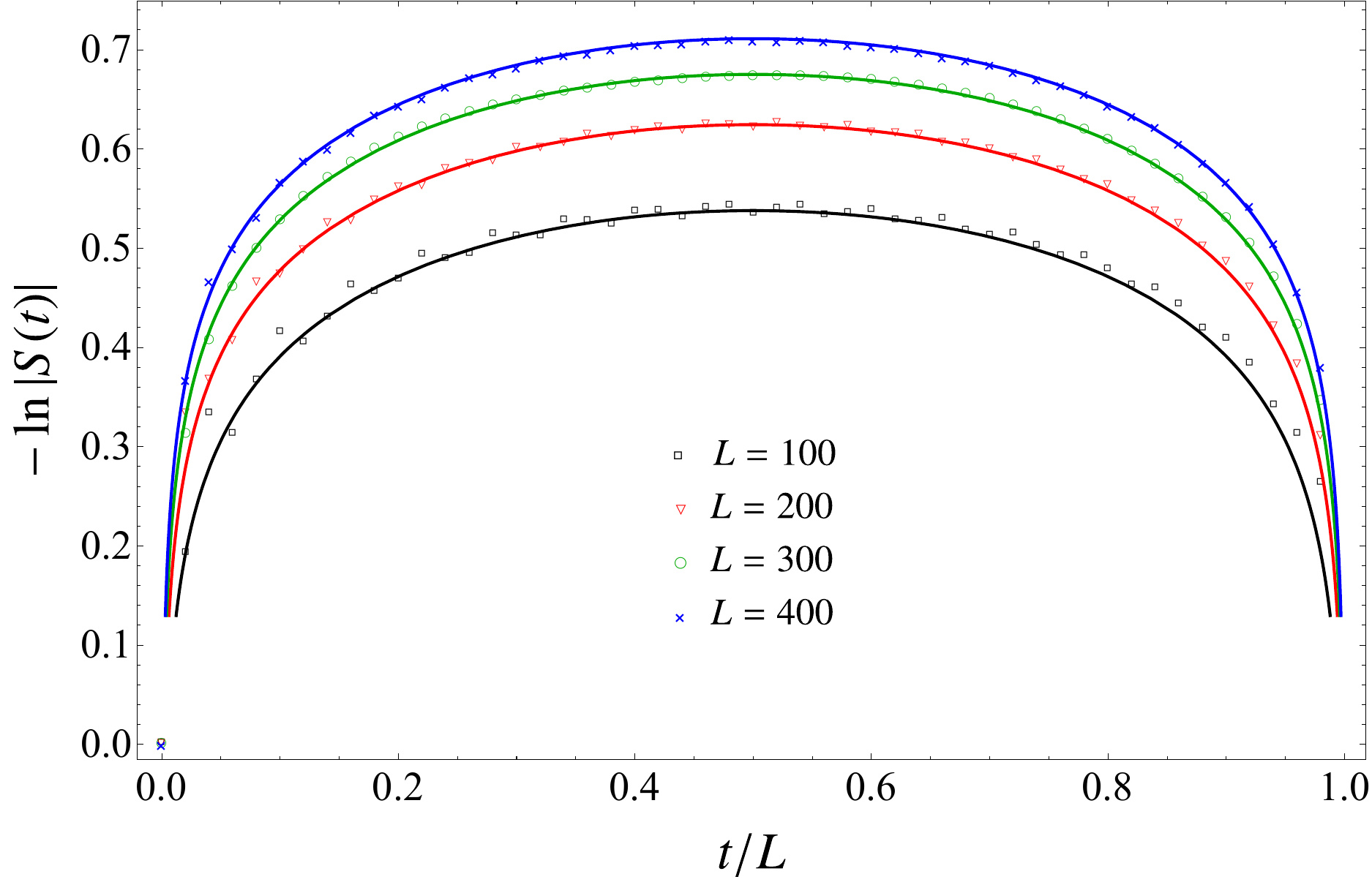}
\hspace{0.85cm}
\includegraphics[width=.45\textwidth]{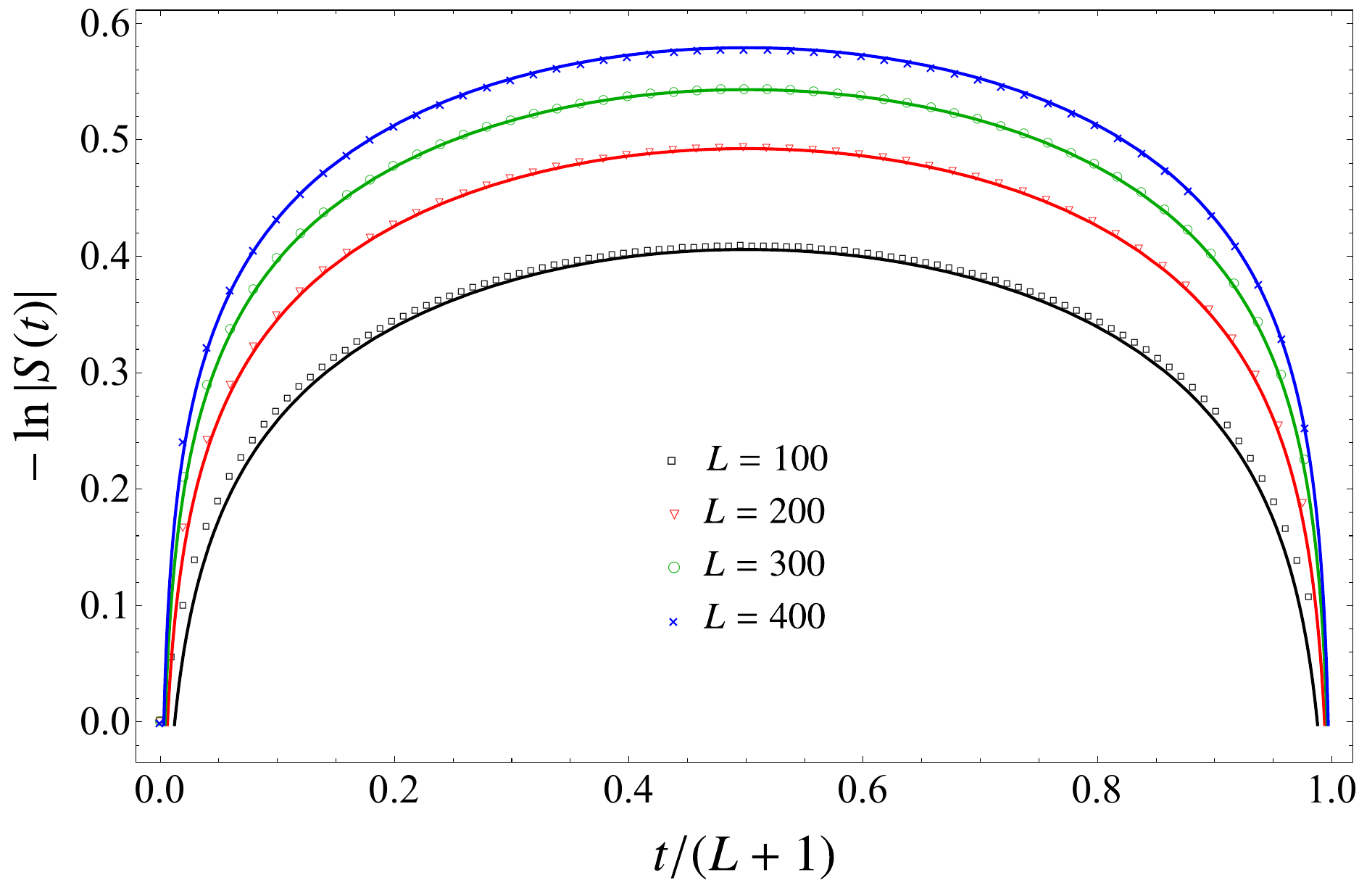}
\caption{
We show the absolute value of the return amplitude after a joining quench of free quantum chains. In both cases, the initial state is a product of identical ground states of the model with $L/2$ sites. The system is let evolve for $t>0$ with the same Hamiltonian, now defined on $L$ sites.
The data are reported as a function of time for various choices of $L$. The solid curves correspond to the CFT predictions (\ref{eq:Snormalized}) with $c=1$ and the additive constant obtained through a fit of $\varepsilon$, resulting in different values for the two models. The points in the left panel are obtained for the tight-binding model with Hamiltonian \eqref{eq:FFHamiltonian} and correspond to a fitted value $\varepsilon=0.215$. In the right panel, the curves are obtained considering a harmonic chain with Hamiltonian \eqref{eq:HC Hamiltonian}, Dirichlet boundary conditions, and a vanishing frequency. In this case, the fit gives $\varepsilon=0.620$.
}
\label{fig:localquenchFreeChains}
\end{figure}

\paragraph{Numerical check of the CFT return amplitudes}
\,
\vspace{.2cm}
\\
In the regime $t\gg2\varepsilon$ where the CFT prediction \eqref{eq:Snormalized} is reliable, the dependence on the UV cutoff occurs only through a multiplicative factor. The cutoff $\varepsilon$ contains non-universal contributions to the return amplitudes. To probe this property, we compute the return amplitude
after the joining quench of two different lattice models that, in the continuum limit, are described by CFTs with the same central charge $c=1$. These are the tight-binding model and a chain of coupled harmonic oscillators (harmonic chain) with vanishing frequency. The evolution of the absolute value of the return amplitude in these lattice models after that two ground states are joined at $t=0$ can be determined using Gaussian techniques. We review this approach in Appendix \ref{app:lattice} while, here, we show and comment on the results.

In Fig.\,\ref{fig:localquenchFreeChains}, we show the evolution of the absolute value of the return amplitudes after a joining quench in two different free quantum chains.
The joining quench occurs by initially preparing the system into the product of two identical ground states of the model with $L/2$ sites and then letting the state evolve with the same Hamiltonian defined on $L$ sites. The curves are reported as a function of time for different values of $L$.
In the left panel, the points are obtained for the tight-binding model while, in the right panel, for the harmonic chain with Dirichlet boundary conditions and zero frequency.
The Hamiltonians of these two models are given in \eqref{eq:FFHamiltonian} and \eqref{eq:HC Hamiltonian} respectively, and their main features are reviewed in Appendices \ref{subapx:FF} and \ref{subapx:HC}.
In the continuum limit, both the models are described by CFTs with $c=1$.
Thus, for large enough values of $L$, we expect a good
agreement with the CFT prediction (\ref{eq:Snormalized}) in the regime $2\varepsilon\ll t\ll L- 2\varepsilon$. This is observed in both panels, where the CFT predictions are represented through solid curves. Since we plot the logarithm of $\vert S(t) \vert$, in the range $2\varepsilon\ll t\ll L- 2\varepsilon$, the only contribution from $\varepsilon$ is in an additive constant. This constant is fitted for the two models leading to $\varepsilon=0.215$ (left) and $\varepsilon=0.620$ (right). The fact that these parameters differ confirms that the denominator in (\ref{eq:Snormalized}) is non-universal. In Sec.\,\ref{sec:Krylovspacequantities}, we will comment on the implications of this fact on Krylov-space quantities.

\subsubsection{Splitting quench}
\label{subsec:returnampl_splitting}
The method discussed so far can also be applied to compute the return amplitude after the splitting quench, which, to our knowledge, has not been done before.
The system defined on an interval of length $L$ is prepared in the ground state $\vert \rm GS \rangle$ of a CFT with Hamiltonian $H_L$ and central charge $c$. At time $t=0$, the system is suddenly divided into two identical parts, each with length $L/2$, and let evolve with the same CFT Hamiltonian $H_{L/2}$ on each of the two parties. More precisely, the evolution Hamiltonian is written as $H=H_{L/2}\otimes \boldsymbol{1}_B+\boldsymbol{1}_A\otimes H_{L/2}$.

\begin{figure}[t!]
\centering
\includegraphics[width=1.\textwidth]{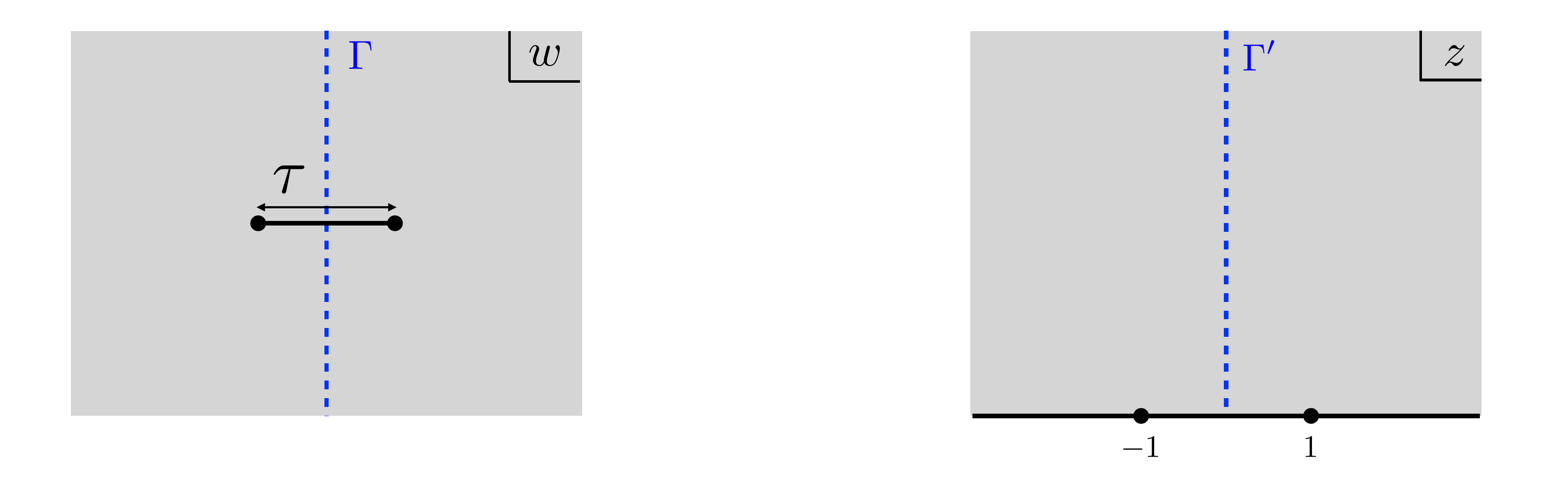}
\caption{In the left panel, we show the two-dimensional geometry whose CFT partition function gives the return amplitude after a splitting quench of an infinite system into two semi-infinite systems. This geometry can be mapped to the half-plane shown in the right panel. The curves $\Gamma$ and $\Gamma'$ used to compute the free energy \eqref{eq:variationF_splitting_v1} are represented as blue dashed lines.}
\label{fig:SplittingQuenchGeom}
\end{figure}

In our computation, we consider $L\to\infty$, or, in other words, the initial ground state of an infinite CFT broken into two semi-infinite lines at $t=0$.
In writing the Euclidean return amplitude $S(\tau)$, we observe that the evolution Hamiltonian $H$, which keeps the two intervals $A$ and $B$ separated, acts for a finite time $\tau$. This suggests that $S(\tau)$ is proportional to the partition function $Z(\tau)$ of the geometry in the left panel of Fig.\,\ref{fig:SplittingQuenchGeom}. This geometry is consistent with the one exploited in \cite{Shimaji:2018czt} for computing the evolution of the entanglement entropy after a splitting quench.

We exploit again \eqref{eq:variationFreeenergy} to compute the free energy \eqref{eq:freeenergydef} associated with $Z(\tau)$. The change of coordinates for mapping the geometry in the left panel of Fig.\,\ref{fig:SplittingQuenchGeom} to the half-plane (right panel) is  
\begin{equation}
\label{eq:mapUHP_splitting}
    w(z)=\tau\frac{z}{z^2+1}\,.
\end{equation}
Using this mapping in \eqref{eq:variationF_Schwarz} and \eqref{eq:defSchwarzian}, we obtain the variation
\begin{equation}
\label{eq:variationF_splitting_v1}
\delta\mathcal{F}=-\frac{c\delta\tau}{4\pi\tau}\int_{\Gamma'} d z \frac{(1+z^2)^2}{(z^2-1)^3} + {\rm c.c.}\,.
\end{equation}
The poles are located at $z=\pm 1$, and we can repeat the computation in Sec.\,\ref{subsec:S_joiningquench} to evaluate the integral. We find
\begin{equation}
\label{eq:variationF_splitting_v1}
\delta\mathcal{F}=\frac{c\delta\tau}{8\pi {\rm i}\tau} 2 \pi {\rm i}\,{\rm Res}_{z=1}\left[\frac{(1+z^2)^2}{(z^2-1)^3}\right] + {\rm c.c.}=\frac{c\delta\tau}{8\tau}
\,.
\end{equation}
This expression is the same as \eqref{eq:variationF_joining_infinite} and, therefore, leads to the same partition function as in \eqref{eq:Paritition_joining_infinite}. This could be argued from the fact that the geometry in the left panel of Fig.\,\ref{fig:SplittingQuenchGeom}, which has a defect on the interval $[-\tau/2,\tau/2]$ along the real axis, can be conformally mapped into the left panel of Fig.\,\ref{fig:JoiningQuenchGeom}, where the defect is in the intervals $(\infty,-\tau/2]\cup[\tau/2,\infty)$. Thus, we expected the partition functions of the same CFT on the two geometries to be the same.   
Physically, this means that the return amplitude after the splitting quench is the same as after the joining quench and is given by \eqref{eq:Snormalized_infinite}. Later, we discuss the implication of this fact on the dynamics of Krylov-space quantities.

As explained in Sec.\,\ref{subsec:S_joiningquench}, this method for computing the return amplitudes relies on the expression of the expectation value of the stress-energy tensor on the quench geometry. Thus, two CFT processes with the same stress-energy tensor lead to the same return amplitude. To study the splitting quench of a thermal CFT state on an infinite line, the geometry in the left panel of Fig.\,\ref{fig:SplittingQuenchGeom} is mapped to a cylinder with a slit of length $\tau$ along the compact direction.
This occurs through the transformation $w\to \frac{1}{2}\tanh\left(\frac{\pi w}{\beta}\right) $, where $\beta$ is the length of the compact dimension of the cylinder and the inverse temperature of the thermal state. 
We can compute the stress-energy tensor in this geometry, finding the same expression as on a cylinder where the defect is located on the intervals $(\infty,-\tau/2]\cup[\tau/2,\infty)$ along the compact direction. This last geometry is used to compute the return amplitude after a joining quench of two CFT thermal states on semi-infinite lines (see Sec.\,\ref{subsec:S_joiningquench} and Appendix \ref{subapp:otherquenches}). We conclude that the return amplitudes after a joining and a splitting quench of thermal states in CFT on infinite lines are identical and given by \eqref{eq:SfiniteT}. By substituting $\beta\to {\rm i} L$, the same result can be exploited for the splitting quench of a CFT ground state on a circle with a single splitting point. The return amplitude is the same as in \eqref{eq:Snormalized}.

\begin{figure}[t!]
\centering
\includegraphics[width=.6\textwidth]{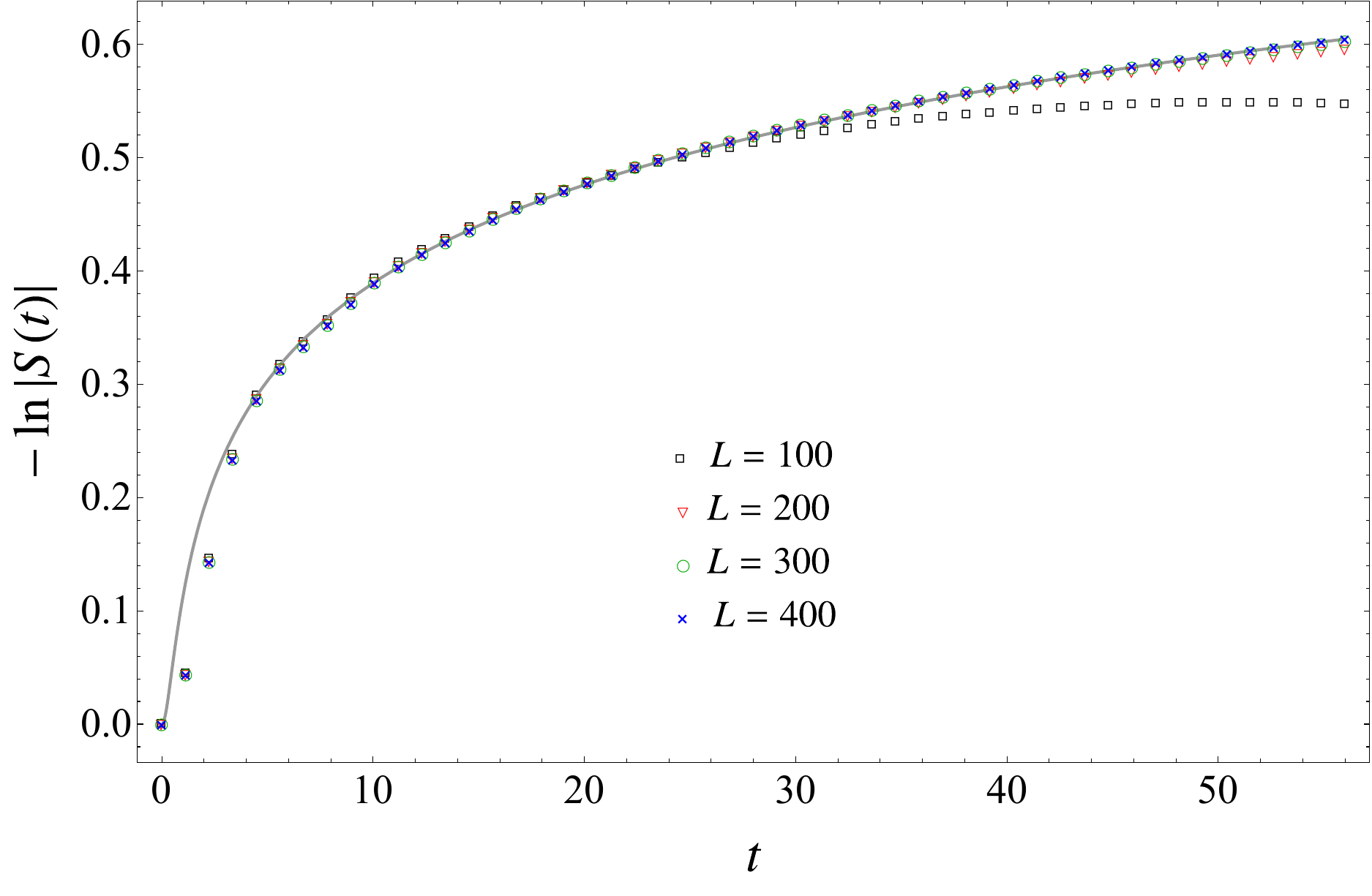}
\caption{The absolute value of the return amplitude after a splitting quench in the tight-binding model with Hamiltonian \eqref{eq:FFHamiltonian}. The initial state is the ground state of the chain $L$ sites. The system is let evolve for $t>0$ with the sum of the Hamiltonian of the model defined on the first $L/2$ sites and the Hamiltonian of the chain on the last $L/2$ sites.
The data are reported as a function of time for various choices of $L$. The solid curve corresponds to the CFT predictions (\ref{eq:Snormalized_infinite}) with $c=1$ and the additive constant obtained through a fit of $\varepsilon=0.225$. We focus on the initial times after the splitting quench in the chain to compare the return amplitude with the CFT predictions obtained for the splitting of a system on an infinite line. }
\label{fig:SplittingReturnAmpl}
\end{figure}

\paragraph{Numerical check of the CFT return amplitudes}
\,
\vspace{.2cm}
\\
In Fig.\,\ref{fig:SplittingReturnAmpl}, we compare the CFT prediction (\ref{eq:Snormalized_infinite}) with the absolute value of the return amplitude after a splitting quench in a tight-binding model. The numerical data are obtained as explained in Appendix \ref{subapp:SplittingFF}. While the CFT result is valid for the splitting quench of an infinite system, we can access only fermionic chains with a finite number of sites $L$. For this reason, we expect agreement between numerical results and analytical prediction only for values of time $t\ll L$. This is indeed observed in Fig.\,\ref{fig:SplittingReturnAmpl}, where we notice that, for smaller values of $L$, the finite-size effects and the consequent deviation from the CFT curve arise earlier. To match the CFT formula with the data points as seen in the figure, we have fitted the parameter $\varepsilon$ in (\ref{eq:Snormalized_infinite}), finding $\varepsilon=0.225$. Interestingly, the value of this parameter is close to the one estimated for the joining quench of the same model (see left panel of Fig.\,\ref{fig:localquenchFreeChains}). The identification of these two UV cutoffs cannot be established due to the lack of an exact lattice computation, which provides a challenging problem for future works.
\subsubsection{Operator quench}
We also provide the expressions of the return amplitudes after the local operator quench introduced in Sec.\,\ref{subsec:operatorquench}.
From the initial state \eqref{eq:initialstate_localopquench}, we notice that determining the return amplitude after the local operator quench in CFT amounts to computing a ratio of two-point functions, namely \cite{Caputa:2023vyr}
\begin{equation}
\label{eq:def_S_opquench}
    S(t)=\frac{\langle{\rm GS}\vert O(x_0,t-{\rm i\varepsilon})O(x_0,{\rm i\varepsilon})\vert{\rm GS}\rangle}{\langle{\rm GS}\vert O(x_0,-{\rm i\varepsilon})O(x_0,{\rm i\varepsilon})\vert{\rm GS}\rangle}\,.
\end{equation}
Since in the CFT setup considered here $\vert\rm GS\rangle$ is the CFT vacuum state and $O$ is a primary operator, the return amplitude in \eqref{eq:def_S_opquench} is fixed by conformal symmetry.
Here, we report the result for the case where the CFT is defined on a circle with circumference $L$. It reads
\begin{equation}
\label{eq:S_opquench_finitesize}
    S(t)=\left[\frac{\sin\left(\frac{\pi (t+2{\rm i}\varepsilon)}{L}\right)}{\sin\left(\frac{2\pi {\rm i}\varepsilon}{L}\right)}\right]^{-2\Delta}
 \,.
\end{equation}
Another insightful initial state we can consider is a thermal CFT state on an infinite line with an operator inserted at a spatial point. In this case, the return amplitude is defined as in \eqref{eq:def_S_opquench} but with the expectation value on the vacuum state replaced by the thermal average. As already discussed in Sec.\,\ref{subsec:S_joiningquench}, if the temperature of the thermal state is $\beta^{-1}$, the expression of the return amplitude is obtained by replacing  $L\to {\rm i}\beta$ in \eqref{eq:S_opquench_finitesize}. The resulting return amplitude is
\begin{equation}
\label{eq:S_opquench_finitetemp}
    S(t)= \left[\frac{\sinh\left(\frac{\pi (t+2{\rm i}\varepsilon)}{\beta}\right)}{\sinh\left(\frac{2\pi {\rm i}\varepsilon}{\beta}\right)}\right]^{-2\Delta}
  \,.
\end{equation}
We conclude this review by pointing at similarities between the return amplitudes after a local operator quench and the corresponding ones after a joining quench. Indeed, we observe that, by replacing $\Delta\to c/16$, \eqref{eq:S_opquench_finitesize} and \eqref{eq:S_opquench_finitetemp} coincide with \eqref{eq:Snormalized} and \eqref{eq:SfiniteT} respectively. We will comment on this feature later concerning the analysis of Krylov-space quantities. 
\subsection{Entanglement entropy after local quenches in CFT}
\label{subsec:Ententropy}
In this section, we review the known results on the evolution of the entanglement entropy after the local quenches described in Sec.\,\ref{subsec:protocols}. This overview is helpful in view of a comparison between the entanglement dynamics and the evolution of Krylov-space quantities after local quenches.

Entanglement entropy is perhaps the most successful measure of quantum correlations. In the last few decades, it has brought many insights into many-body physics, especially when studied in out-of-equilibrium settings. As an entanglement measure, the entanglement entropy necessitates a spatial bipartition of the system considered. In the local quench protocols reviewed in Sec.\,\ref{subsec:protocols}, the evolving system is always naturally divided into two complementary parts, either by a defect (joining and splitting quench) or by an operator insertion (operator quench). Thus, it is meaningful to consider the entanglement entropy of one of these parts, say the left one. Given the state $\vert\psi(t)\rangle$ evolved at time $t$, the entanglement entropy of the left part of the system is defined as 
\begin{equation}
\label{eq:EEdefinition}
 S_{\rm left}(t)=-{\rm Tr}\left[\rho_{\rm left}(t)\ln\rho_{\rm left}(t)\right]   \,,
    \qquad
    \rho_{\rm left}(t)\equiv{\rm Tr}_{\rm right}\left[\vert\psi(t)\rangle\langle\psi(t)\vert\right] 
    \,,
\end{equation}
where the reduced density matrix $\rho_{\rm left}(t)$ is obtained by tracing out the degrees of freedom in the right part of the system.
For later convenience, here we report the expressions of the entanglement entropy of the left half of the system along the local quench dynamics described in Sec.\,\ref{subsec:protocols}. In this manuscript, we do not report any derivation of these results, referring the interested reader to the works where these are found (for a comprehensive introduction on entanglement entropy in CFT, see \cite{Calabrese:2009qy}).

If the initially decoupled CFT ground states are defined on equal intervals of length $L/2$, the evolution of the entanglement entropy of the left part after the joining quench reads \cite{Stephan:2011kcw}
\begin{equation}
\label{eq:EElocalquench finite}
S_{\rm left}(t)=\frac{c}{3}\ln\left\vert\frac{L}{\pi \varepsilon}\sin\left(\frac{\pi t}{L}\right)\right\vert+{\rm const}\,,
\end{equation}
where the additive constant contains non-universal contributions.
In the limit $L\to\infty$, i.e. when the two intervals joined at $t=0$ become semi-infinite lines, \eqref{eq:EElocalquench finite} reduces to \cite{Calabrese:2007mtj}
\begin{equation}
\label{eq:EElocalquench infinite}
S_{\rm left}(t)=\frac{c}{3}\ln(t/\varepsilon)+{\rm const}\,.
\end{equation}
Interestingly, the prefactor of the logarithmic growth of the entanglement entropy depends only on the central charge of the underlying CFT. Thus, this evolution allows to extract universal properties of the systems under consideration.

In the case of the splitting quench, the result is qualitatively different. Indeed, for the splitting quench of a CFT of an infinite line into two semi-infinite lines, the entanglement entropy is constant in time \cite{Shimaji:2018czt}. This can be explained by observing that, despite the left and the right parts being initially entangled, after the splitting, the entanglement cannot increase since the entangling point coincides with the breaking point. Notice that the constant behavior of the entanglement entropy is a feature of this choice of bipartition in an infinite system. The time evolution of the entanglement of other bipartitions is non-trivial, as discussed in \cite{Shimaji:2018czt}.

Finally, we turn to the local operator quench. Considering the setup of Sec.\,\ref{subsec:operatorquench} in the limit $L\to\infty$ of infinite system, we focus on the dynamics of the entanglement entropy of the left half-infinite line starting at the insertion point $x_0$.
As shown in \cite{Nozaki:2014hna}, the evolution of the entanglement after this quench can be computed by exploiting the replica trick and computing first the Rényi entropies defined as
\begin{equation}
 S^{(n)}_{\rm left}(t)=\frac{1}{1-n}\ln{\rm Tr}\left[\rho^n_{\rm left}(t)\right]   \,,
\end{equation}
and then taking the analytic continuation in $n$ and the limit $n\to 1$ to obtain the entanglement entropy \eqref{eq:EEdefinition}. The Rényi entropies are determined by inserting two operators $O$ on each replica and computing $2n$-point correlation functions of these operators on the replica manifold. The $2n$-point correlation functions of primary operators in CFT are not fixed by conformal invariance, and their expression depends on the specific CFT considered as well as on the operator $O$. This implies that the evolution of the Rényi entropies and the entanglement entropy after this protocol is less universal and explicitly depends on the CFT and the operator inserted in the initial state. This remark will be crucial in comparing entanglement entropy and K-entropy, as reported in Sec.\,\ref{subsec:Kentropy}.
To give an example, if we considered (1+1)-dimensional Rational CFTs, the entanglement growth with respect to the initial state is given by the logarithm of the quantum dimension of $O$ \cite{Nozaki:2014hna}
\begin{equation}
\label{eq:localopquench_EE}
  \Delta S^{(n)}_{\rm left}(t)=\ln d_O  \,,
\end{equation}
and is valid also for the entanglement entropy when $n=1$.
On the other hand, for large c holographic CFTs, we can derive the result for the von Neumann entropy even at finite temperature using the vacuum block approximation \cite{Caputa:2014vaa,Caputa:2015waa} 
\be
\Delta S^{(1)}_{\rm left}(t)=\frac{c}{6}\log\left(\frac{\beta}{\pi\varepsilon}\sinh\left(\frac{\pi t}{\beta}\right)\frac{\sinh(\pi\alpha_O)}{\alpha_O}\right),\qquad \alpha_O=\sqrt{\frac{24h_O}{c}-1}\,.
\ee
Interestingly, if we naively substitute $h_O\to c/32$ \footnote{Which is in fact quite subtle \cite{Kusuki:2017upd,Kusuki:2019gjs} and generally the results for Renyi and von Neumann entropies depend on the operator's dimension.} we end up with
\be
\Delta S^{(1)}_{\rm left}(t)=\frac{c}{6}\log\left(\frac{2\beta}{\pi\epsilon}\sinh\left(\frac{\pi t}{\beta}\right)\right)\simeq \frac{c}{6}\log\left(\frac{2t}{\varepsilon}\right)+O(1/\beta^2)\,.
\ee
This is again a logarithmic growth with time but with a factor of 2 difference with respect to the more universal, joining and splitting quenches.
\section{Complexity and entropy in the Krylov basis}
\label{sec:Krylovspacequantities}
This section presents the main findings of our work, highlighting the time evolution of spread complexity and K-entropy following various types of local quenches.
\subsection{Effective dynamics with SL(2,$\mathbb{R}$) symmetry}
To make our discussion self-contained, we begin with a brief review of the computation of spread complexity and K-entropy when the dynamics on the Krylov chain is governed by an effective or emergent SL(2,$\mathbb{R}$) symmetry. Here, we summarize only the key formulas relevant to our forthcoming analysis, while a more detailed discussion can be found in \cite{Caputa:2021sib, Balasubramanian:2022tpr,Dymarsky:2019elm}.

Let us consider an auxiliary problem of the real time evolution through a Hamiltonian written in terms of the SL(2,$\mathbb{R}$) generators
\begin{equation}
\label{eq:Hevolution_SL2R}
    H_{\rm SL(2,\mathbb{R})}=\gamma L_0+\alpha(L_1+L_{-1})+\delta\,,
\end{equation}
where the parameters $\alpha$, $\gamma$ and $\delta$ are real numbers related to some physical properties and the SL(2,$\mathbb{R}$) generators satisfy the algebra
\begin{equation}
  [L_n,L_m]=(n-m)L_{m+n}  \,,
    \qquad
   n,m=-1,0,1 \,.
\end{equation}
As usual, $L_{-1}$ plays the role of the raising and $L_{1}$ of the lowering ladder operator.\\
We assume that the initial state is the highest weight state $\vert h\rangle$ specified by
\begin{equation}
  L_0\vert h\rangle=h\vert h\rangle  \,,
    \qquad
  L_1\vert h\rangle=0 \,.
\end{equation}
Then, we can follow the evolution in the natural, orthonormal basis of the Lie algebra defined as
\begin{equation}
   \vert h,n\rangle\equiv\sqrt{\frac{\Gamma(2h)}{n!\Gamma(2h+n)}} L^n_{-1}\vert h\rangle \,,
    \qquad
   \langle h,n \vert h,m\rangle=\delta_{nm}\,.\label{AlgBasis}
\end{equation}
Observe that, using the action of the algebra generators on the basis vectors
\begin{eqnarray}
  L_{0} \vert h,n\rangle&=&(h+n)\vert h,n\rangle \,,\nn
   \\
  L_{1} \vert h,n\rangle&=&\sqrt{n(n+2h-1)}\vert h,n-1\rangle \,,\nn
   \\
   L_{-1}\vert h,n\rangle&=&\sqrt{(n+1)(n+2h)}\vert h,n+1\rangle \,,\label{ActionLs}
\end{eqnarray}
we find that the Hamiltonian \eqref{eq:Hevolution_SL2R} is tri-diagonal in this basis. In fact, the algebra basis \eqref{AlgBasis} is the Krylov basis for this Hamiltonian, and we can automatically read-off Lanczos coefficients from \eqref{ActionLs}.
\begin{equation}
\label{eq:Lanczoscoefficients}
    a_{n}=\gamma(h+n)+\delta\,,
    \qquad\qquad
    b_{n}=\alpha\sqrt{n(2h+n-1)}\,.
\end{equation}

Choosing the highest weight vector as the initial state allows us to easily expand the evolved state on the Krylov basis. Indeed, from the Baker-Campbell-Hausdorff formula, we find
\begin{equation}
\label{eq:expansiondynamicsKrylovbasisSL2R}
 \vert \psi(t)\rangle=e^{-{\rm i}H_{\rm SL(2,\mathbb{R})}t}   \vert h\rangle=e^{-{\rm i}\delta t}e^{hB}\sum_{n=0}^\infty\sqrt{\frac{\Gamma(2h+n)}{n!\Gamma(2h)}} A^n\vert h,n\rangle\,,
\end{equation}
where
\begin{equation}
    A\equiv-\frac{2 {\rm i} \alpha}{\mathcal{D}}\frac{\tanh{\left(\frac{\mathcal{D}t}{2}\right)}}{1+\frac{ {\rm i} \gamma}{\mathcal{D}}\tanh{\left(\frac{\mathcal{D}t}{2}\right)}}
\,,
\qquad
e^{-B}\equiv\left[\cosh\left(\frac{\mathcal{D}t}{2}\right)+\frac{ {\rm i} \gamma}{\mathcal{D}}\sinh\left(\frac{\mathcal{D}t}{2}\right)\right]^2
\,,
\end{equation}
and 
\begin{equation}
\label{eq:mathcalDdef}
    \mathcal{D}\equiv\sqrt{4\alpha^2-\gamma^2}\,.
\end{equation}
The expansion \eqref{eq:expansiondynamicsKrylovbasisSL2R} provides the analytical expressions for the amplitudes introduced in \eqref{eq:Krylov expansion}, which read
\begin{equation}
\label{eq:amplKSpace}
  \psi_n(t)= e^{-{\rm i}\delta t}\sqrt{\frac{\Gamma(2h+n)}{n!\Gamma(2h)}} e^{hB}A^n \,.
\end{equation}
Remarkably, the probabilities associated with \eqref{eq:amplKSpace} can be rewritten in terms of a unique function of the parameters in \eqref{eq:Hevolution_SL2R} as
\begin{equation}
\label{eq:probKSpace}
    p_n(t)=\vert\psi_n(t)\vert^2=\frac{\Gamma(2h+n)}{n! \Gamma(2h)}X^n (1-X)^{2h}\,,
\end{equation}
where
\begin{equation}
\label{eq:Xdef}
   X=\frac{\sinh^2\left(\frac{\mathcal{D} t}{2}\right)}{\cosh^2\left(\frac{\mathcal{D} t}{2}\right)-\frac{\gamma^2}{4 \alpha^2}}\in [0,1]\,,
\end{equation}
and $\mathcal{D}$ is defined in \eqref{eq:mathcalDdef}.
It is straightforward to check by summing over the index $n$ that $p_n(t)$ defines a normalized probability. Moreover, from \eqref{eq:amplKSpace}, we can extract the return amplitude along this evolution as
\begin{equation}
\label{eq:returnamplitude_SL2R}
    S(t)=\psi^*_0(t)=e^{{\rm i}\delta t}\left[\cosh\left(\frac{\mathcal{D}t}{2}\right)-\frac{ {\rm i} \gamma}{\mathcal{D}}\sinh\left(\frac{\mathcal{D}t}{2}\right)\right]^{-2h}\,.
\end{equation}
Since the dynamics in the Krylov space is completely determined by the return amplitude, if we have an evolution where $S(t)$ can written as in \eqref{eq:returnamplitude_SL2R}, we can describe it on the Krylov chain in terms of an effective SL(2,$\mathbb{R}$) Hamiltonian \eqref{eq:Hevolution_SL2R}, where the parameters $\alpha$, $\gamma$, and $\delta$ are written in terms of the physical quantities involved in the process. 

Once the dynamics has been rewritten in the Krylov basis, we can use the probabilities \label{eq:probKSpace} to compute the spread complexity and the K-entropy.
Applying the definition \eqref{eq:spread complexity} and using the probabilities \eqref{eq:probKSpace}, we find the spread complexity
\begin{equation}
\label{eq:spreadcomplexitySL(2R)}
C_\mathcal{K}(t)=\frac{8 h\alpha^2}{\mathcal{D}^2}\sinh^2{\left(\frac{\mathcal{D} t}{2}\right)}\,.
\end{equation}
The parameter $\mathcal{D}$ can be real or purely imaginary based on the sign of $4\alpha^2-\gamma^2$. This is crucial to establish the time behaviour of $C_\mathcal{K}(t)$. Indeed, for real $\mathcal{D}$, $C_\mathcal{K}(t)$ grows exponentially at late times, while, if it is purely imaginary, the spread complexity has an oscillatory behaviour. The intermediate regime with vanishing  $\mathcal{D}$ corresponds to quadratic growth of spread complexity.

As for the K-entropy, the sum in \eqref{eq:Kentropy_def} cannot be generally written in a closed form if the probabilities are given by \eqref{eq:probKSpace}.  
Only for $h=1/2$, the sum can be performed analytically, leading to
\begin{equation}
\label{eq:H_honehalf}
   H_{\mathcal K}=\frac{-X\ln X-(1-X)\ln(1-X)}{1-X}\,.
\end{equation}
More generally, we can reduce the full sum to the expression 
\begin{eqnarray}
\label{eq:H_anyh}
   H_{\mathcal K}&=&2h\frac{-X\ln X-(1-X)\ln(1-X)}{1-X}+\ln\Gamma(2h)
   \\
  &-&\frac{(1-X)^{2h}}{\Gamma(2h)}\sum_{n=0}^\infty \frac{\Gamma(2h+n)}{n!}X^n\ln\left(\frac{\Gamma(2h+n)}{n!}\right)\,,
  \nonumber
\end{eqnarray}
valid for any value of $h$.
For later convenience, we find it worth stressing that, in the regime $X\to 1$, the leading terms are the ones shown in the first line of \eqref{eq:H_anyh}. This consideration will be helpful in Sec.\,\ref{subsec:Kentropy} when we study the late time regime of the K-entropy in explicit examples.

\subsection{Spread complexity after local quenches}
\label{subsec:spreadComplexity}
Let us now compute the evolution of the spread complexity during local quench scenarios described in Sec.\,\ref{subsec:protocols}. The key observation for this purpose is that the return amplitudes that we have derived in Sec.\,\ref{subsec:returnampl_CFT} can be written in the SL(2,$\mathbb{R}$) form  \eqref{eq:returnamplitude_SL2R}.

We begin our analysis from the joining quench of two CFT ground states defined on finite systems of length $L/2$. The return amplitude along this dynamics is given by \eqref{eq:Snormalized} and can be written as \eqref{eq:returnamplitude_SL2R} if we fix the parameters in \eqref{eq:Hevolution_SL2R} as
\begin{equation}
\label{eq:parametersSL2R}
    \alpha =\frac{\pi}{L\sinh\frac{2\pi\varepsilon}{L}}\,,
    \qquad
    \gamma=\frac{2\pi}{L}\coth\frac{2\pi \varepsilon}{L}\,,
    \qquad
    h=\frac{c}{16}\,,
    \qquad
     \delta=-\frac{\pi c}{6 L}\,.
\end{equation}
This implies that the joining quench evolution can be effectively described by a dynamics on the Krylov chain with an emergent SL(2,$\mathbb{R}$) symmetry. Interestingly, the dependence on the central charge $c$ enters into the representation weight $h$. We will discuss this more at the end of this section.

Consequently, the spread complexity of this process can be computed directly from \eqref{eq:spreadcomplexitySL(2R)} by using \eqref{eq:parametersSL2R}. We find
\begin{equation}
\label{eq:spreadcomplexity}
    C_{\mathcal K}(t)=\frac{c}{8}\left(\frac{\sin\frac{\pi t}{L}}{\sinh\frac{2\pi \varepsilon}{L}}\right)^2
    \simeq
    \frac{c}{32}\frac{L^2 \sin^2\frac{\pi t}{L}}{\pi^2\varepsilon^2}
    \,.
\end{equation}
The spread complexity has a periodic time dependence with a period equal to $L$ and depends on all the parameters of the CFT quench dynamics, including the central charge $c$. Before commenting on this intriguing feature, we discuss the spread complexity of the local quench dynamics in other regimes of interest and for other protocols considered in this work. Indeed, in Sec.\,\ref{subsec:S_joiningquench}, we mentioned the regime $L\to\infty$, corresponding to the joining quench of two semi-infinite lines. In this limit, the SL(2,$\mathbb{R}$) parameters \eqref{eq:parametersSL2R} become 
\begin{equation}
\label{eq:parametersSL2R_infinite}
    \alpha =\frac{1}{2\varepsilon}\,,
    \qquad
    \gamma=\frac{2}{2\varepsilon}\,,
    \qquad
    h=\frac{c}{16}
    \,,
    \qquad
     \delta=0\,,
\end{equation}
and, since they corresponds to $\mathcal{D}=0$, they lead to the spread complexity
\begin{equation}
\label{eq:spreadcomplexitylargeL}
    C_{\mathcal K}(t)=\frac{c}{32}\frac{t^2}{\varepsilon^2}\,.
\end{equation}
Clearly, the multiplicative dependence on the central charge is preserved in this limit.

In Sec.\,\ref{subsec:S_joiningquench} we also considered two semi-infinite CFTs at finite temperature $\beta^{-1}$ joined at $t=0$. Comparing the return amplitude \eqref{eq:SfiniteT} with \eqref{eq:returnamplitude_SL2R}, we fix the SL(2,$\mathbb{R}$) parameters as
  \begin{equation}
    \label{eq:parametersSL2R_finiteTemp}
         \alpha =\frac{\pi}{\beta\sin\frac{2\pi\varepsilon}{\beta}}\,,
    \qquad
    \gamma=\frac{2\pi}{\beta}\cot\frac{2\pi \varepsilon}{\beta}\,,
    \qquad
    h=\frac{c}{16}\,,
    \qquad
     \delta=0\,.
    \end{equation}
    This leads to the following spread complexity 
    \begin{equation}
    \label{eq:spreadcomplexity_finiteTemp}
       C_{\mathcal K}(t)=\frac{c}{8}\left(\frac{\sinh\frac{\pi t}{\beta}}{\sin\frac{2\pi \varepsilon}{\beta}}\right)^2
       \simeq
    \frac{c}{32}\frac{\beta^2 \sinh^2\frac{\pi t}{\beta}}{\pi^2\varepsilon^2}\,.
    \end{equation}
Differently from the oscillatory behaviour in \eqref{eq:spreadcomplexity} and the quadratic growth in \eqref{eq:spreadcomplexitylargeL}, the spread complexity \eqref{eq:spreadcomplexity_finiteTemp} grows exponentially in time.

The growth of the spread complexity after the splitting quench can be inferred from the results reported above. Indeed, in Sec.\,\ref{subsec:returnampl_splitting}, we noticed that the return amplitude of the splitting quench of a CFT on an infinite line into two semi-infinite lines is given by \eqref{eq:Snormalized_infinite} and is the same as the joining quench of two semi-infinite CFT ground states. Since the dynamics on the Krylov chain is determined by the return amplitude only, the spread complexity after the splitting quench is the same as after the joining quench, and its expression is \eqref{eq:spreadcomplexitylargeL}.
The same holds for the spread complexity after a splitting quench of an infinite CFT thermal state and a CFT ground state on a circle, given by \eqref{eq:spreadcomplexity_finiteTemp} and \eqref{eq:spreadcomplexity} respectively.
Thus, all these spread complexities have the same features discussed for the joining quench, including the multiplicative dependence on the central charge. 

The dependence of the spread complexity on $c$ is probably the most intriguing feature found in our analysis so far. Indeed, to the best of our knowledge, a spread complexity explicitly dependent on the central charge was never observed. The central charge identifies the universality class of the system. Thus, the dependence of $C_{\mathcal K}(t)$ on $c$ implies that the spread complexity is sensitive to this physical property. However, from all the expressions reported in this section, we observe that the central charge always appears multiplied by $\varepsilon^{-2}$. As we have discussed and verified in Sec.\,\ref{subsec:S_joiningquench}, the UV cut-off contains non-universal information on the system. This means that, despite the dependence of the spread complexity on the universality class of the model, a purely universal prefactor cannot be extracted from the its time evolution. In the following section, we will address a similar question for the time evolution of the K-entropy.

For completeness, we report the expression for the spread complexity after a CFT local operator quench, where the initial state is locally excited by a primary operator. This result was originally derived in \cite{Caputa:2024sux}. As in the previous cases, we can map the return amplitude \eqref{eq:S_opquench_finitesize} of this process to \eqref{eq:returnamplitude_SL2R}, obtaining in this way $\alpha$, $\gamma$, $h$ and $\delta$ as functions of the operator quench parameters. Plugging these functions into \eqref{eq:spreadcomplexitySL(2R)}, we obtain
\begin{equation}
\label{eq:spreadcomplexity_localoperator}
    C_{\mathcal K}(t)=2\Delta\left(\frac{\sin\frac{\pi t}{L}}{\sinh\frac{2\pi \varepsilon}{L}}\right)^2
    \simeq
   \frac{\Delta L^2 }{2\pi^2\varepsilon^2}\sin^2\frac{\pi t}{L}
    \,.
\end{equation}
As commented in Sec.\,\ref{subsec:operatorquench} for the return amplitudes, this result can be alternatively obtained by substituting $c/16 \to \Delta$ in \eqref{eq:spread complexity}. Finally, the spread complexity after a local operator quench in a thermal CFT on an infinite line can be obtained from \eqref{eq:spreadcomplexity_localoperator} by replacing $L\to {\rm i} \beta$.

We conclude our analysis of spread complexity by emphasizing that the effective description of dynamics on the Krylov chain, governed by the Hamiltonian \eqref{eq:Hevolution_SL2R} with an emergent SL(2,$\mathbb{R}$) symmetry, is not a universal feature of all local quench protocols. Instead, it crucially depends on the choice of the initial state. Indeed, in Appendix \ref{app:domainwallquench}, we examine an alternative local quench protocol, the domain wall quench, where the initial state differs from the ones considered in the main text. In particular, the states defined on the initially separated left and right parts are different. The evolution of the return amplitude after the domain wall quench does not have the form \eqref{eq:returnamplitude_SL2R}, but we can still identify a different emergent symmetry for this dynamics and solve it analytically. 
\subsection{K-entropy after local quenches}
\label{subsec:Kentropy}
Next, we discuss the time evolution of K-entropies after the local quenches described in Sec.\,\ref{subsec:protocols} and compare it with the evolution of von Neumann entanglement entropies. Indeed, we can compute the K-entropies using \eqref{eq:H_anyh} and the fact that all the local quenches discussed in this work can be effectively described on the Krylov chain with an SL(2,$\mathbb{R}$) symmetry. Thus, for all our examples, the function $X$ in \eqref{eq:Xdef} can be written in terms of the parameters of the CFT quenches exploiting their mapping into the parameters of the Hamiltonian \eqref{eq:Hevolution_SL2R}. In what follows, we focus on the joining quench, but the same results also hold for the K-entropies after the splitting quench since the dynamics on the Krylov chain are the same for the two processes.

In \eqref{eq:parametersSL2R}, we report the relations between the SL(2,$\mathbb{R}$) parameters and the physical quantities that determine a joining quench dynamics, where the initial state is made of two disconnected CFT ground states on equal intervals of length $L/2$. Plugging them into \eqref{eq:Xdef}, we find 
\begin{equation}
\label{eq:X_finite}
X=\frac{\sin^2\left(\frac{\pi t}{L}\right)}{\sin^2\left(\frac{\pi t}{L}\right)+\cosh^2\left(\frac{2\pi \varepsilon}{L}\right)-1}\,,
\end{equation}
and, inserting this expression into \eqref{eq:H_anyh} with $h=c/16$, we obtain the K-entropy for the joining quench protocol.
The expression \eqref{eq:X_finite} implies that the K-entropy is periodic with period $L$. 
\begin{figure}[t!]
\hspace{0.15cm}
\includegraphics[width=.45\textwidth]{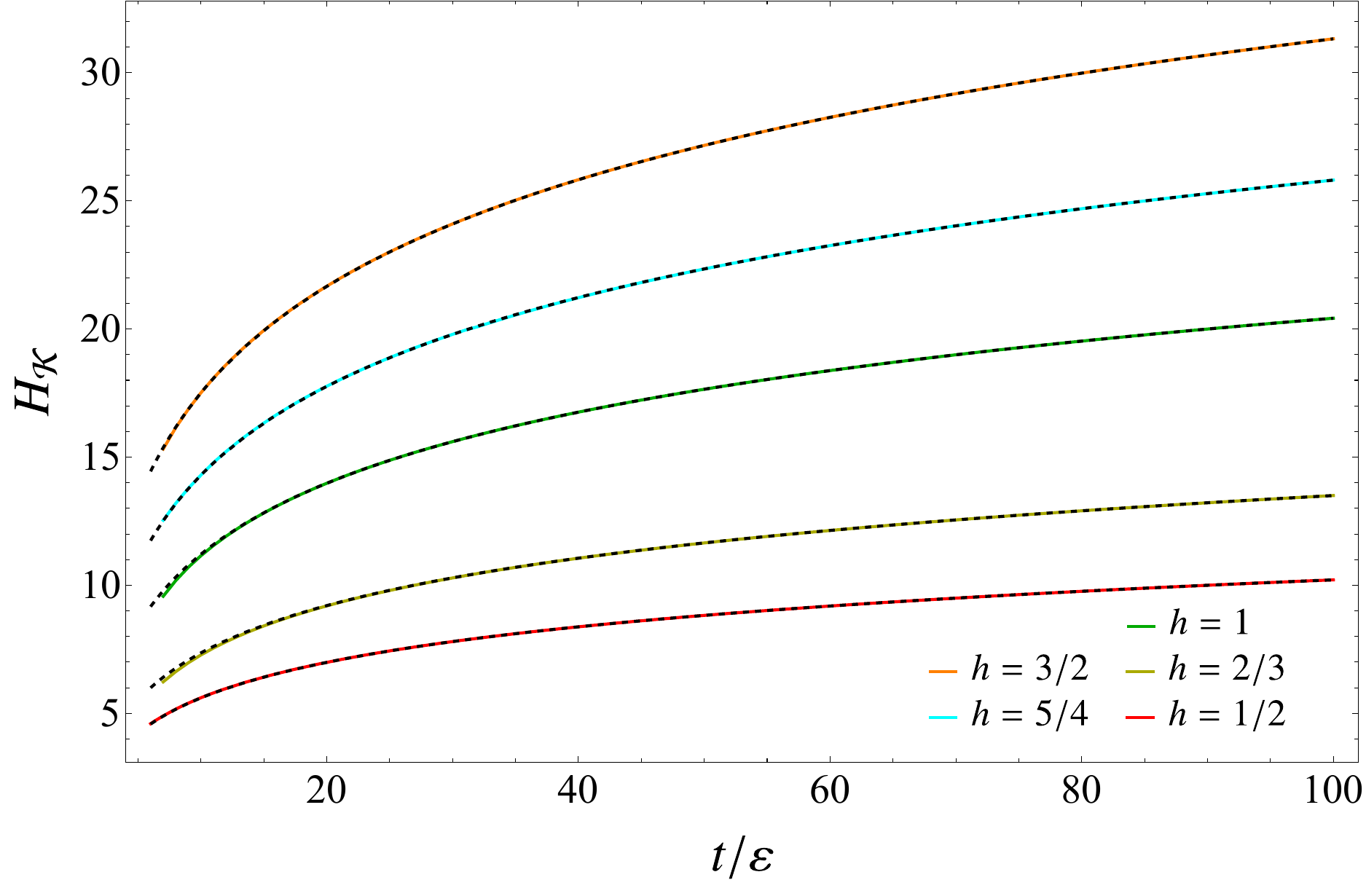}
\hspace{0.85cm}
\includegraphics[width=.45\textwidth]{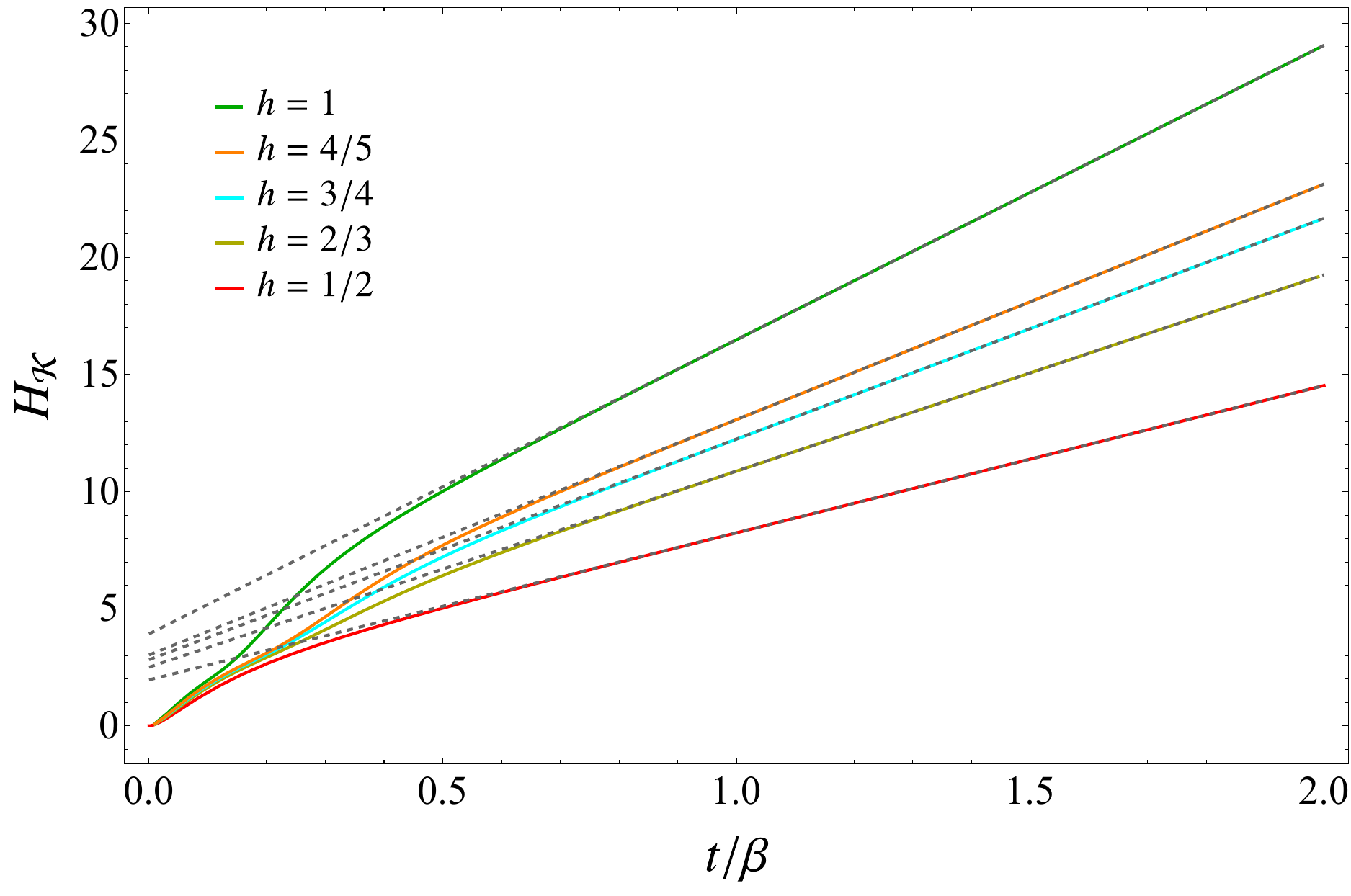}
\caption{
K-entropies obtained from \eqref{eq:H_honehalf} and \eqref{eq:H_anyh} plotted as a function of a dimensionless time parameter for different values of $h=c/16$. In the left panel, we consider a joining quench of two semi-infinite CFT ground states. In this case, the function $X$ is given by \eqref{eq:X_ininite}. The dashed black curves are obtained from \eqref{eq:H_infinite_larget}. The K-entropies in the right panel are obtained for a joining quench of two-semi infinite CFTs at finite temperature $\beta^{-1}$. The expression of $X$ is given by \eqref{eq:X_finiteTemp}, and the UV cutoff is chosen so that $\varepsilon/\beta=0.05$. The dashed gray lines are obtained from \eqref{eq:H_finiteTemp_larget}. 
}
\label{fig:Kentropies}
\end{figure}
We find it insightful to consider the special case of the joining quench of two semi-infinite CFT ground states. This is achieved by taking the limit $L\to\infty$. In this regime, the parameter $X$ becomes 
\begin{equation}
\label{eq:X_ininite}
X=\frac{t^2/\varepsilon^2}{4+t^2/\varepsilon^2}\,.
\end{equation}
As already mentioned, the sum in \eqref{eq:H_anyh} cannot be computed for general values of the parameters, but we can gain analytical insights by studying the late time regime. Using \eqref{eq:X_ininite} in \eqref{eq:H_anyh} and taking the limit $t^2/(4\varepsilon^2)\to\infty$, i.e. $X\to 1$, we obtain
\begin{equation}
\label{eq:H_infinite_larget}
   H_{\mathcal K}= \frac{c}{8} \big[1+2\ln \left(t/(2\varepsilon)\right)\big]+\ln\Gamma(c/8)+ O(t^{-2})\,.
\end{equation}
At late times, the K-entropy exhibits a logarithmic growth with a prefactor proportional to the central charge only, without non-universal contributions. This asymptotic behavior is represented by gray dashed curves in the left panel of Fig.\,\ref{fig:Kentropies} compared to the full K-entropies \eqref{eq:H_anyh} (colored curves) obtained for $X$ in \eqref{eq:X_ininite} and $h=c/16$.

A late-time growth of the K-entropy with a prefactor proportional to the central charge is also obtained by considering a joining quench of two thermal CFT states on semi-infinite lines. By using \eqref{eq:parametersSL2R_finiteTemp} in \eqref{eq:Xdef}, we obtain 
\begin{equation}
\label{eq:X_finiteTemp}
X=\frac{\sinh^2\left(\frac{\pi t}{\beta}\right)}{\sinh^2\left(\frac{\pi t}{\beta}\right)+1-\cos^2\left(\frac{2\pi \varepsilon}{\beta}\right)}\,,
\end{equation}
which provides the K-entropy once plugged into \eqref{eq:H_anyh} with $h=c/16$. At late times $t/\beta \gg 1$, the K-entropy reads
\begin{equation}
\label{eq:H_finiteTemp_larget}
   H_{\mathcal K}= \frac{c}{8} \left(1+2\pi \frac{t}{\beta}-\ln\left[4 \sin^2(2\pi \varepsilon/\beta)\right]\right)+\ln\Gamma(c/8)+ O\left(e^{-2\pi t/\beta}\right)\,.
\end{equation}
In this case, the late time growth is linear, and the slope depends only on the temperature and the central charge, without any non-universal parameters.
This behaviour has been reported in the right panel of Fig.\,\ref{fig:Kentropies} (dashed gray lines) where we also show the K-entropies obtained from the full sum \eqref{eq:H_anyh} with $X$ given by \eqref{eq:X_finiteTemp} (colored curves) and $h=c/16$.


Interestingly, the K-entropies \eqref{eq:H_infinite_larget} and \eqref{eq:H_finiteTemp_larget} after joining (and splitting) quenches of infinite CFTs detect universal properties of the system. Indeed, by looking at the time growth, we can read-off from the prefactor the central charge of the CFT, i.e. the universality class to which the considered system belongs. This feature is not found for the spread complexities \eqref{eq:spreadcomplexitylargeL} and \eqref{eq:spreadcomplexity_finiteTemp}, where we observed a prefactor dependent on the central charge, but such a dependence occurs with a multiplicative non-universal UV cutoff. Thus, we conclude that the evolution of the K-entropies after joining and splitting quenches is more appropriate than the spread complexities for detecting universal properties.

The last quench protocol we analyze from the point of view of the K-entropy evolution is the local operator quench described in Sec.\,\ref{subsec:operatorquench}. For the evolution from a CFT ground state locally excited by a primary operator with scaling dimension $\Delta$, we exploit the fact that the return amplitude (and therefore the full dynamics on the Krylov chain) is the same as the joining quench up to the replacement $c/16 \to \Delta$. Thus, performing this substitution in $X$ in \eqref{eq:X_finite}, we can compute the K-entropy after a CFT local operator quench. In the limit of large system size, the late-time growth reads
\begin{equation}
\label{eq:H_infinite_operator_larget}
   H_{\mathcal K}= 2\Delta \big[1+2\ln (t/(2\varepsilon))\big]+\ln\Gamma(2\Delta)+ O(t^{-2})\,,
\end{equation}
exhibiting the same logarithmic growth as the joining and splitting quenches but no dependence on the central charge.

It is also worth mentioning that there have been definitions and studies of entropies obtained by projecting a (reduced) density matrix onto a given basis, generally different from the Krylov basis. This problem was addressed in \cite{Alcaraz:2013ara,Alcaraz:2014tfa,Stephan:2014nda} for states projected on eigenvectors of spin operators, leveraging CFT techniques. When considering the projection of a density matrix reduced to an interval of finite length $\ell$, intriguing similarities arise between the results in \cite{Alcaraz:2013ara} and the K-entropies discussed in this section (up to the exchange $t\leftrightarrow\ell$). Apart from using the same CFT tools outlined in Sec.\,\ref{subsec:returnampl_CFT}, it would be interesting to understand more about possible physical properties shared between the two cases.

To conclude our analysis of the K-entropies, we compare our findings with the time evolution of the half-space entanglement entropy after local quenches, reviewed in Sec.\,\ref{subsec:Ententropy}. The K-entropy is a notion of entropy associated with the dynamic of the entire system.
However, differently from the entanglement entropy, it does not require a bipartition of the system and does not directly compute the entanglement between complementary parties. For this reason, there is no a priori reason to expect any similarity between these two quantities. Despite this fact, the analysis of the different local quench protocols considered in this work leads to interesting observations. 

First, we remark that, since the entanglement entropies depend on the considered bipartition, we have to choose one for quantitative comparison. A natural choice is the bipartition into the left and right halves that are initially separated (by boundaries or operator insertions) and then joined at $t=0$. Thus, we compare $H_{\mathcal K}$ computed in the various cases in this section with the corresponding $S_{\rm left}(t)$, whose expressions are reported in Sec.\,\ref{subsec:Ententropy}.
For the local quench where we join two CFT ground states on semi-infinite lines, the left entanglement entropy is given by
\eqref{eq:EElocalquench infinite}.
This result can be compared with the late time behavior \eqref{eq:H_infinite_larget}. The two expressions share the same logarithmic growth in time, but the prefactor of the leading terms of the entanglement entropy is $4/3$ times the one of the Krylov space entropy.
The situation changes for the splitting quench of an infinite CFT into two halves. The late-time growth of the Krylov space entropy is again \eqref{eq:H_infinite_larget}, but the left entanglement entropy is constant, as discussed in Sec.\,\ref{subsec:Ententropy}. Thus, in the splitting quench, the entanglement and the Krylov space entropy have a distinct behavior in time.
Finally, the difference between the time evolution of the K-entropy \eqref{eq:H_infinite_operator_larget} and the entanglement entropy \eqref{eq:localopquench_EE} after a local operator quench is even sharper. Indeed, we recall that the entanglement entropy evolution in this scenario depends on the CFT we consider and the primary excitation inserted in the initial state. On the contrary, the logarithmic growth in \eqref{eq:H_infinite_operator_larget} is the same for any CFT with a prefactor uniquely determined by the scaling dimension of the inserted operator. In summary, we have found that the time evolution of the half-space entanglement entropy and the K-entropy after local quenches is generally different. However, after the local joining quench, the two entropies exhibit the same logarithmic growth, with prefactors differing rescaled by $4/3$. It would be interesting to understand whether this similarity is accidental or arises from a more profound reason.
\subsection{Comments on Lanczos coefficients and large c}
Last but not least, we comment on the presence of the central charge in our formulas and differences between small and large central charge limits that play important roles in CFTs that appear in holography.

Observe that our single quench scenarios generally lead to the following Lanczos coefficients governed by the SL(2,$\mathbb{R}$) symmetry
\be
a_n=\gamma\left(n+\frac{c}{16}\right),\qquad b_n=\alpha\sqrt{n\left(\frac{c}{8}+n-1\right)}\,.
\ee
This is interesting because the central charge introduces an effective scale for the Krylov index $n$, i.e., the position on the Krylov chain. Indeed, if $0\le n\ll c$,  we can first extract the large c limit on the level of the Lanczos coefficients
\bea
a_n&=&\frac{c}{16}\gamma\left(1+\frac{16n}{c}\right)\simeq \frac{c}{16}\gamma+O(1/c)\,,\nn\\
 b_n&=&\alpha\sqrt{\frac{c}{8}n\left(1+\frac{8(n-1)}{c}\right)}\simeq \alpha\sqrt{\frac{c}{8}}\sqrt{n}+O(1/\sqrt{c})\,.
\eea
The solutions of the Schrodinger equation \eqref{eq:Schroedinger eq_amplitude} with these coefficients fall into the Heisenberg-Weyl family \cite{Caputa:2021sib} 
\be
\psi_n(t)=\frac{(-{\rm i}\sqrt{c}\,\alpha t)^n}{2^{\frac{3n}{2}}\sqrt{n!}}e^{-\frac{c}{16}t\left(t\alpha^2+i\gamma\right)}\,,
\ee
and give rise to the spread complexity that grows quadratically with time
\be
C_{\mathcal{K}}(t)=\frac{c}{8}\alpha^2t^2\,.
\ee
If we compare this with the full answer, we can see that this is equivalent to taking the early time limit. This is also consistent with the fact that large c bounds $n$ to the times where semi-classical gravity provides a good description. If $n$ is larger and/or comparable to $c$ we cannot do this approximation anymore and the full SL(2,$\mathbb{R}$) structure with the exponential growth is important. Of course, because we are describing the CFT dynamics with infinite dimensional Hilbert space, we will not be able to see the saturation, such as e.g. in random matrix models \cite{Balasubramanian:2022tpr}.
\section{Local quenches in holography and spread complexity}\label{sec:Holography}
In this section, we describe the holographic dual of local quenches and discuss how to extract the spread complexity of the quench states considered above. Since a progress in this direction was already done for local operator quenches in \cite{Caputa:2024sux}, we start from this setup.
\subsection{Point Particle description}
Local operator quenches can be modeled holographically using massive point particles in AdS spacetimes \cite{Nozaki:2013wia,Caputa:2014vaa}. In fact, this approach was first developed as a model for local quenches in holography. 

More precisely, one considers asymptotically $AdS_3$ spacetime with metric $g_{\mu \nu}(x)$ and a massive point particle following trajectory $x^\mu(\lambda)$. Particle's action is given by
\begin{equation}
  S_\text{m} = -m\int d\lambda\,\sqrt{-g_{\mu \nu}(x)\dot{x}^\mu\dot{x}^\nu}\,,
\label{eq:pa}
\end{equation}
and its extremum yields the geodesic's trajectory. For example, in Poincare coordinates
\be
ds^2=\frac{-dt^2+dx^2+dz^2}{z^2}\,,
\ee
the trajectory is given by
\be
\lambda=t,\qquad x=0,\qquad z=\sqrt{t^2+\varepsilon^2}\,,\label{PPTrajectory}
\ee
where at $t=0$ the particle approaches $z=\varepsilon$ that matches the operator's regulator in CFT. One can further take the back-reaction from the point particle by solving Einstein's equations with the particle as a source for the stress-energy tensor. Such geometry with a shock-wave localized on particle's trajectory can then be used to reproduce the evolution of correlation functions or entanglement entropy computed using the HRT prescription \cite{Hubeny:2007xt}.

Recently, using this holographic setup and motivated by \cite{Susskind:2014jwa,Susskind:2018tei,Magan:2018nmu,Lin:2019qwu,Susskind:2020gnl,Barbon:2020uux}, it was found that the rate of growth of spread complexity is proportional to the proper radial momentum of the point particle \cite{Caputa:2024sux}
\begin{equation}
     \partial_tC_{\mathcal K}(t)=-\frac{1}{\varepsilon}P_\rho(t)\,,
\label{eq:CP}
\end{equation}
where the coordinate $\rho$ is the proper radial distance in the bulk (see also \cite{Fan:2024iop,He:2024pox} for related works). For the example above, the proper momentum is measured in coordinate $\rho=-\log(z)$ where the metric becomes
\be
ds^2=d\rho^2+e^{2\rho}(-dt^2+dx^2)\,.
\ee
The proper momentum is computed in the canonical way from the particles Lagrangian\footnote{Defined via $S_m=\int dt \mathcal{L}_m $ .}
\be
P_\rho(t)=\frac{\partial \mathcal{L}_m}{\partial \rho'(t)}\,,
\ee
and the relation to the rate of complexity is obtained by identification of the particle's mass with conformal dimension of the CFT operator $m=\Delta$.

Observe that the bulk description of the local operator quench dynamics is confined to the $x=0$ plane that is the $AdS_2$ slice of the $AdS_3$ geometry. This is quite intuitive and natural, given the fact that dynamics on the Krylov chain is governed by the emergent SL(2,$\mathbb{R}$) symmetry. Nevertheless, the relation between the symmetry generators on the chain and the SL(2,$\mathbb{R}$) isometries of $AdS_2$ in the bulk are non-trivial\footnote{See \cite{Lin:2019qwu} for a similar and perhaps related discussion in lower dimensions that would be interesting to connect to.}. 

Given our simple form of return amplitudes and spread complexities for the local quench scenarios, one could expect that the local quench should also be modeled by a massive particle with mass
\be
m=\frac{c}{16}\,.
\ee
This would agree with the original proposal \cite{Nozaki:2013wia} and confirm the relation \eqref{eq:CP}. However, the holographic description of local quenches has recently been improved/corrected using the AdS/BCFT and we will discuss how to extract the rate of complexity in this framework. 
\subsection{The AdS/BCFT description}
As we saw in the first part of the paper, local quenches in 2d CFTs, such as the joining \cite{Calabrese:2007mtj} and splitting \cite{Shimaji:2018czt} one, are analyzed using  boundary conformal field theory (BCFT). It is then natural to develop their gravity dual using the framework of AdS/BCFT \cite{Takayanagi:2011zk,Fujita:2011fp}. Indeed this was studied for joining and splitting quenches in \cite{Ugajin:2013xxa,Astaneh:2014fga,Shimaji:2018czt}. We will first review this briefly in the following.

The main part of the holographic proposal for a dual of a BCFT \cite{Takayanagi:2011zk} is to consider a portion of AdS spacetime with BCFT sub-region at its boundary and the end-of-the-world (EOW) brane $Q$ in the bulk. To fully specify this portion of the holographic geometry we consider the action
\be
I=\frac{1}{2\kappa^2}\int d^3x\sqrt{-g}(R-2\Lambda)+\frac{1}{\kappa^2}\int_Q d^2x\sqrt{|h|}(K-T)\,,
\ee
where $\kappa^2=8\pi G_N$ and $K$ is the extrinsic curvature on $Q$ and $T$ is its tension\footnote{In the AdS/BCFT construction the action also involves the Gibbons-Hawking term on the cut-off surface $\Sigma$ that we omit since it will not play any role in our discussions.}. The variation of the action yields Einstein's equations as well as the Neumann b.c. on $Q$
\be
K_{ab}-K h_{ab}=-T h_{ab}\,.\label{NBC}
\ee
For holographic local quenches that we discuss here, it was shown \cite{Shimaji:2018czt} that surface $Q$ requires zero tension $T=0$.

The main step is then to find $Q$ that constitutes the boundary of the holographic AdS/BCFT setup. In 3 dimensions, this can be done by using the trick of extending the relevant map used in the CFT, to the $AdS_3$ bulk. More precisely, starting from the Poincare metric
\bea
ds^2=\frac{d\eta^2+d\xi d\bar{\xi}}{\eta^2}\,,
\eea
we can apply the change of coordinates
\bea
&& \xi=f-\frac{2z^2(f')^2(\bar{f}'')}{4f'\bar{f}'+z^2f''\bar{f}''}\,,\nn\\
&& \bar{\xi}=\bar{f}-\frac{2z^2(\bar{f}')^2(f'')}{4f'\bar{f}'+z^2f''\bar{f}''}\,,\nn\\
&& \eta=\frac{4z(f'\bar{f}')^{3/2}}{4f'\bar{f}'+z^2f''\bar{f}''}\,. \label{MapToBañados}
\eea
where $f(w)$ and $\bar{f}(\bar{w})$ (with $w=x+t$ and $\bar{w}=x-t$ in Lorentzian signature) are holomorphic/anti-holomorphic maps. The resulting geometry is described by the so-called Bañados metric \cite{Banados:1998gg}
\be
ds^2=\frac{dz^2}{z^2}+\frac{(dw+z^2\bar{T}(\bar{w})d\bar{w})(d\bar{w}+z^2T(w)dw)}{z^2}\,,\label{BañadosG}
\ee
with 
\be
T(w)=-\frac{1}{2}\{f(w),w\},\qquad \bar{T}(\bar{w})=-\frac{1}{2}\{\bar{f}(\bar{w}),\bar{w}\}\,.
\ee
In AdS/BCFT, functions $f(w)$ and $\bar{f}(\bar{w})$ are taken as maps between the BCFT region and the upper half plane (u.h.p) with coordinates $(\xi,\bar{\xi})$. In the bulk, in $(\eta,\xi,\bar{\xi})$ coordinates, surface $Q$ is specified by the condition $Im(\xi)=0$, i.e. $\xi=\bar{\xi}$.  Therefore, after inserting \eqref{MapToBañados}, we can solve this constraint to find $Q$ specified by $z=z(t,x)$ in the Bañados metric as \cite{Kudler-Flam:2023ahk}\footnote{For compactness of our formulas, we suppress the explicit dependence on the arguments.}
\be
z(t,x)=\sqrt{\frac{4(f-\bar{f})f'\bar{f}'}{2(f'^2\bar{f}''-\bar{f}'^2f'')-(f-\bar{f})f''\bar{f}''}}\,.\label{QMostGen}
\ee
With this EOW surface at hand, the holographic setup is complete and we can e.g. compute the evolution of entanglement entropies using the HRT prescription \cite{Hubeny:2007xt} and its AdS/BCFT counterparts \cite{Takayanagi:2011zk}. Moreover, functions $T(w)$ and $\bar{T}(\bar{w})$ are directly related to the expectation values of the CFT stress tensor in the BCFT (for us the quench) state
\be
\langle \psi(t)| T(w)|\psi(t)\rangle=\frac{c}{12}\{f(w),w\}\,,
\ee
and similarly for $\bar{T}(\bar{w})$. In the following, we will discuss this construction from the perspective of the spread complexity of the local joining quench derived in the first part of the work. 
\subsubsection{Vacuum joining quench}
Using the conventions of \cite{Shimaji:2018czt,Caputa:2019avh}, the geometry describing the single local joining quench of two CFTs at zero temperature is the complex plane with slits along the Euclidean time $(-\infty,-{\rm i}\varepsilon]\cup[{\rm i}\varepsilon,+\infty)$. It can be mapped to the upper half plane with the following function
\be
f(w)={\rm i}\sqrt{\frac{w+{\rm i}\varepsilon}{{\rm i}\varepsilon-w}}\,.
\label{MapQPlane}
\ee
After extending it to the $AdS_3$ bulk with \eqref{MapToBañados}, we end up with spacetime \eqref{BañadosG} with
\be
T(w)=\frac{3\varepsilon^2}{4(w^2+\varepsilon^2)^2}, \qquad \bar{T}(\bar{w})=\frac{3\varepsilon^2}{4(\bar{w}^2+\varepsilon^2)^2}\,.\label{STLocJQ}
\ee
In other words, the expectation values of the stress tensors in CFTs describe two wave packets that propagate away from the joining point at $x=0$ to the left and to the right with the speed of light. The total energy in the quench state is constant and computed as\footnote{We use the convention $T_{00}=-\frac{1}{2\pi}(T+\bar{T})$.}
\be
E=\int dx\langle \psi(t)|T_{00}(x,t)|\psi(t)\rangle =\frac{c}{16\varepsilon}\,,
\ee
and is consistent with the local operator quench result after replacing $m=\Delta=c/16$.

\begin{figure}[t!]
\centering
\includegraphics[width=1\textwidth]{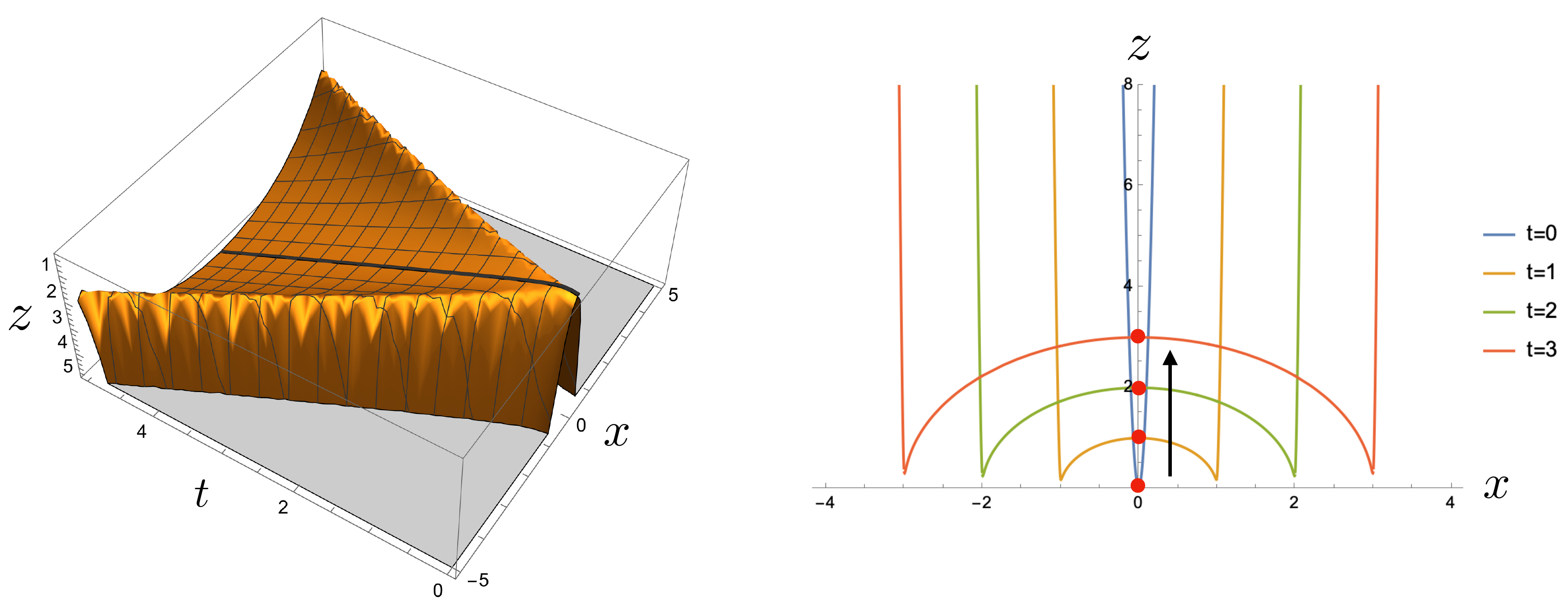}
\caption{Left: 3D Plot of the surface \eqref{eqn:z3D} and its $x=0$ slice is marked in black. Right: time slices of the surface at various instances of time. The tip of the surface (marked with red dots) probes the bulk the deepest and behaves effectively like a point particle in geometry \eqref{BañadosG}.}
\label{fig:AdSBCFT}
\end{figure} 
Using \eqref{QMostGen}, we obtain the parametrization of the surface $Q$ that solves the Neumann b.c. \eqref{NBC} with $T=0$ as \cite{Caputa:2019avh}
\be
z(t,x)=\frac{2\sqrt{x^2+(\varepsilon+{\rm i}t)^2}\sqrt{x^2+(\varepsilon-{\rm i}t)^2}}{\sqrt{\varepsilon^2-\left(\sqrt{x^2+(\varepsilon+{\rm i}t)^2}-\sqrt{x^2+(\varepsilon-{\rm i}t)^2}\right)^2}}\,,\label{eqn:z3D}
\ee
and we present it on Fig. \ref{fig:AdSBCFT}. At $t=0$ the EOW brane is folded along the radial direction at $x=0$. Then it ``opens up" as time progresses, but its portions along the light-cones always remain near the boundary, mimicking the evolution of the energy density in CFT.

Naively, such a time-dependent geometry seems very complicated and it is not obvious how dynamics represented holographically by this spacetime can be captured by SL(2,$\mathbb{R}$) symmetry on the Krylov chain. However, it is interesting to observe that the tip of the EOW brane located at $x=0$ (depicted by red dots on the right of Fig. \ref{fig:AdSBCFT} for several instances of time) is exploring the bulk along a trajectory analogous to the one of a massive point particle\footnote{Similar observations were discussed also in \cite{Mezei:2016wfz}.} \eqref{PPTrajectory} from the previous subsection
\be
z(t,0)=\frac{2(t^2+\varepsilon^2)}{\sqrt{4t^2+\varepsilon^2}}\,.
\ee
The main difference is that at $t=0$ the tip of $Q$ is located at $z=2\varepsilon$, whereas the particle starts from $z=\varepsilon$, but the motion of both objects is confined to the two dimensional $x=0$ plane. It is then tempting to follow \cite{Caputa:2024sux} and examine the proper momentum of the tip of $Q$ as well as its relation with the rate of spread complexity in CFTs. 

Firstly, we can indeed check that trajectory of the tip extremizes the action of a point particle on the $x=0$ slice of geometry \eqref{BañadosG}. We will go straight to the proper coordinate $z=\exp(-\rho)$ such that the action of a particle following trajectory $\rho=F(t)$ becomes
\be
S_m=-m\int dt\sqrt{e^{2F}-T-\bar{T}+T\bar{T}\,e^{-2F}-F'^2}\,,\label{LagBanGen}
\ee
where the explicit time dependence enters via $T(0+t)$ and $\bar{T}(0-t)$ in \eqref{STLocJQ}. The solution of the equations of motion is precisely the trajectory of the tip of $Q$
\be
F(t)=-\log\left(\frac{2(t^2+\varepsilon^2)}{\sqrt{4t^2+\varepsilon^2}}\right)\,.
\ee
The proper momentum can be expressed as
\be
P_\rho(t)=\frac{mF'}{\sqrt{e^{2F}-T-\bar{T}+T\bar{T}\,e^{-2F}-F'^2}}\,,
\ee
and for our solution  becomes
\be
P_\rho(t)=-\frac{2mt}{\varepsilon}=-\frac{c}{8\varepsilon}t\,.\label{PRad1}
\ee
where we inserted $m=c/16$. Comparing with the rate of growth of the spread complexity \eqref{eq:spreadcomplexitylargeL} we find that
\be
\partial_tC_{\mathcal K}(t)=-\frac{1}{2\varepsilon}P_\rho(t)\,.\label{CPinAdSBCFT}
\ee
This is the same relation as \eqref{eq:CP} but with factor of $2$ in front of the $\varepsilon$. Let us first test it for the thermal quench and then we will discuss the possible origin of this factor. 
\subsubsection{Thermal joining quench}
We can perform a similar analysis for the local joining quench at finite temperature studied in \cite{Mezei:2019zyt}. The quench state is now prepared on the cylinder with a compact Euclidean time of period $\beta$ and a slit   $\tau\in [-\beta/2,-\varepsilon]\cup[\varepsilon,\beta/2]$ at $x=0$, representing the joining of two semi-infinite halves at finite temperature at $\tau=-{\rm i}\varepsilon$. This geometry, with coordinate $w=x+{\rm i}\tau$, is then mapped to the upper half plane by
\be
\xi=f(w)={\rm i}\sqrt{\frac{\sinh\left(\frac{\pi(w+{\rm i}\varepsilon)}{\beta}\right)}{\sinh\left(\frac{\pi({\rm i}\varepsilon-w)}{\beta}\right)}}\,,\label{MapsFinTemp}
\ee
and similarly for $\bar{\xi}=\bar{f}(\bar{w})$. This is simply the composition of the map from the cylinder with the slit to the plane
\be
g(w)=\frac{1}{2}\tanh\left(\frac{\pi w}{\beta}\right)\,,
\ee
that we had in the zero temperature example\footnote{Remembering that $g(w)$ maps the slits from $\varepsilon$ on the cylinder to those starting at $\frac{1}{2}\tan\left(\frac{\pi\varepsilon}{\beta}\right)$ on the plane.}
followed by \eqref{MapQPlane}.

The expectation value of the stress tensor in CFT is again consistent with the local operator quench \cite{Caputa:2014eta}, after we set the chiral dimension to $h=c/32$ ($\Delta=2h$)
\be
\langle\psi(t)| T(w)|\psi(t)\rangle=\frac{h\pi^2}{\beta^2}\frac{\sinh^2\left(\frac{2\pi {\rm i}\varepsilon}{\beta}\right)}{\sinh^2\left(\frac{\pi (w-{\rm i}\varepsilon)}{\beta}\right)\sinh^2\left(\frac{\pi (w+{\rm i}\varepsilon)}{\beta}\right)}-\frac{c\pi^2}{6\beta^2},\qquad h=\frac{c}{32}\,,
\ee
and similarly for $\bar{T}(\bar{w})$.

The parametrization \eqref{QMostGen} of the surface $Q$ is now much more complicated but its $x=0$ slice, i.e. the tip, can be written as
\be
z(t,0)=\frac{2\sqrt{2}\beta\left(\cosh\left(\frac{2\pi t}{\beta}\right)-\cos\left(\frac{2\pi \varepsilon}{\beta}\right)\right)}{\pi\sqrt{4\cosh\left(\frac{2\pi t}{\beta}\right)-3-\cos\left(\frac{4\pi\varepsilon}{\beta}\right)}}\,.
\ee
At $t=0$ the tip approaches the boundary to the distance 
\be
z(0,0)=\frac{2\beta}{\pi}\tan\left(\frac{\pi\varepsilon}{\beta}\right)\sim 2\varepsilon\,,
\ee
and for late times it reaches up to the radial distance
\be
\lim_{t\to\infty}z(t,0)=\frac{\beta}{\pi}\,.
\ee
To compute the radial momentum, we again start with \eqref{BañadosG} with functions $T(w)$ and $\bar{T}(\bar{w})$ proportional to Schwarzian derivatives of \eqref{MapsFinTemp}, and introduce the radial coordinate $z=\beta/\pi\, e^{-\rho}$. The trajectory of the tip of $Q$ is again determined from Lagrangian of the analogous functional form as \eqref{LagBanGen} but with different functions of time $t$. The solution becomes
\be
F(t)=-\log\left(\frac{\pi}{\beta}z(t,0)\right)\,,
\ee
such that the radial momentum for $m=c/16$ becomes
\be
P_\rho(t)=-\frac{c\beta}{16\pi\varepsilon}\sinh\left(\frac{2\pi t}{\beta}\right)\,.\label{PRad2}
\ee
Comparing with the rate of \eqref{eq:spreadcomplexity_finiteTemp} we again get relation \eqref{CPinAdSBCFT} between the rate of spread complexity and the proper radial momentum of the tip of $Q$ in AdS/BCFT setups.
Analogous results for finite size can be obtained by sending $\beta\to {\rm i}L$ and the AdS/BCFT setup has been studied in detail in \cite{Kudler-Flam:2023ahk} for this case.

A couple of comments are in order. We have already seen that standard entanglement entropies for local joining and operator quenches had a factor of 2 difference so it is not surprising to find similar pattern in our relations between the rate of complexity and proper momenta in these two different scenarios. Nevertheless, in order to gain more intuition behind this factor of $2$ we can compare our result above with the trajectory of the point particle that describes local operator quench at finite temperature using the BTZ metric
\be
ds^2=\frac{1}{\tilde{z}^2}\left(-(1-M\tilde{z}^2)dt^2+\frac{d\tilde{z}^2}{1-M\tilde{z}^2}+dx^2\right),\quad M=\frac{4\pi^2}{\beta^2}\,.
\ee
It is given by
\be
\tilde{z}(t)=\frac{\beta}{2\pi}\sqrt{1-\left(1-\frac{4\pi^2\varepsilon^2}{\beta^2}\right)\left(1-\tanh^2\left(\frac{2\pi t}{\beta}\right)\right)}\,.
\ee
At $t=0$, the particle starts at $\tilde{z}=\varepsilon$ and for $t\to\infty$ approaches the horizon at $\tilde{z}=1/\sqrt{M}=\beta/2\pi$. The relation between the BTZ radial coordinate and the Bañados metric (that is also in the Fefferman-Graham gauge) is
\be
\tilde{z}=\frac{z}{1+z^2T(w)}\,,
\ee
where the $T(w)=\bar{T}(\bar{w})=\pi^2/\beta^2=-\frac{1}{2}\{f(w),w\}$ where $f(w)=\exp(2\pi w/\beta)$ is the map from the thermal cylinder to the plane. 

Then indeed the horizon in the $z$ coordinates corresponds to $z=\beta/\pi$, so half of the distance in BTZ and that is where the tip of $Q$ asymptotes to at late times. Moreover, the radial coordinate
\be
z=\frac{\beta}{\pi}e^{-\rho}\,,
\ee
is the proper distance to the horizon and gives the metric used in \cite{Caputa:2024sux}
\be
ds^2=d\rho^2+\frac{4\pi^2}{\beta^2}\left(-\sinh^2(\rho)dt^2+\cosh^2(\rho)dx^2\right)\,.
\ee
It is tempting to expect that this factor of 2 between BTZ and Fefferman-Graham coordinates from the AdS/BCFT setup might be the main suspect of the 2 in \eqref{CPinAdSBCFT}. Unfortunately this is not the case and even if we rescale the coordinate $z\to 2\hat{z}$ such that the $Q$ starts at $\hat{z}=\varepsilon$ and for late times goes to $\hat{z}=\beta/2\pi$, the radial momentum is still the same as \eqref{PRad2} (and the same for \eqref{PRad2}). 
On the other hand, holographically, we may expect that factor of 2 for entanglement entropy in operator vs local joining quench can be related to the difference between the standard HRT prescription and the connected vs disconnected contributions in AdS/BCFT. Understanding the connection between these prescriptions for computing entanglement (as well as the membrane picture for its evolution in chaotic systems \cite{Mezei:2018jco,Jonay:2018yei}) and the Krylov approach is beyond the scope of this work but is among the most interesting future directions.

In any case, it is very reassuring that the effective motion of the tip of $Q$ in the 2D plane describes the rate of spread complexity so well. It will be very interesting to explore this relation further for other local single and double quenches as well as more complicated setups where the SL(2,$\mathbb{R}$) symmetry does not govern the Krylov chain anymore, and we hope to report on it in the future.
\section{Conclusions and discussion}\label{sec:Conclusions}
In this work, we revisit local quench dynamics in two-dimensional CFTs using the Krylov space approach. We have focus on determining Lanczos coefficients, spread complexity, and K-entropy after different types of local quenches. These include the joining quench, where two initially disconnected parts are suddenly joined and evolve as a single system, and the splitting quench, where a homogeneous system is divided into two separately evolving parts. Moreover, we extend existing Krylov space results on the local operator quench, where the initial state is a CFT ground state locally excited by a primary operator.    

The dynamics of the Krylov chain are analyzed through the return amplitude. In Sec.\,\ref{subsec:S_joiningquench}, we review the CFT methods from \cite{Stephan:2011kcw} to compute this quantity after a joining quench, both for finite and infinite system. In Sec.\,\ref{subsec:returnampl_splitting}, we generalize this approach to the splitting quench of CFT states, deriving the return amplitudes \eqref{eq:Snormalized} (finite system), \eqref{eq:Snormalized_infinite} (infinite system) and \eqref{eq:SfiniteT} (finite temperature), which, to our knowledge, are presented here for the first time. Interestingly, the return amplitude for a splitting quench in an infinite system mirrors that of a joining quench for two CFT states on semi-infinite lines. 

Crucially, all the return amplitudes computed in Sec.\,\ref{subsec:returnampl_CFT} can be rewritten in the form of return amplitudes associated with an effective SL(2,$\mathbb{R}$)-symmetric evolution. This allows us to derive analytical results for the spread complexity and the K-entropy across all types of local quenches considered in this work. 

For a local quench involving two CFT ground states on finite intervals, the spread complexity is given by \eqref{eq:spreadcomplexity}. In the limit where the initially separated intervals become semi-infinite, it simplifies to \eqref{eq:spreadcomplexitylargeL}. Similarly, for a joining quench of two semi-infinite CFTs in a thermal state at equal temperature, the spread complexity is given by \eqref{eq:spreadcomplexity_finiteTemp}.
For a splitting quench of an infinite CFT ground state, the spread complexity again takes the form of \eqref{eq:spreadcomplexitylargeL}. In contrast, for a splitting quench of a finite system or a system at finite temperature, the spread complexities are given by \eqref{eq:spreadcomplexity} and \eqref{eq:spreadcomplexity_finiteTemp}, respectively.

Remarkably, all the spread complexities derived in Sec.\,\ref{subsec:spreadComplexity} have a prefactor, which is proportional to the central charge of the CFT, encoding information about the system’s universality class. However, this prefactor also exhibits an algebraic dependence on the UV cutoff, preventing a clear separation between universal and non-universal contributions in the spread complexity expressions. 
In contrast, the evolution of Krylov entropies reveals a prefactor that depends on the central charge but is independent of the non-universal cutoff. This feature can be seen from the late-time expressions \eqref{eq:H_infinite_larget} (joining and splitting quench of infinite systems) and \eqref{eq:H_finiteTemp_larget} (joining and splitting quench of infinite thermal states). This suggests that the K-entropies might be more appropriate for capturing the universal features of local quench dynamics. 

We find it insightful to compare the evolution of K-entropies with that of half-space entanglement entropies. Interestingly, for a joining quench, the two entropies exhibit similar time evolution, differing only by a numerical prefactor that is independent of the quench details. However, for other quench protocols, their behavior differs qualitatively. In particular, comparing these two quantities for local operator quenches, we find that the K-entropy is universal for all CFT and depends only on the scaling dimension of the inserted operator (see \eqref{eq:H_infinite_operator_larget}). On the other hand, the functional form of the corresponding entanglement entropy varies depending on the CFT under consideration.

Finally, in Sec.\,\ref{sec:Holography}, we explore the spread complexity of joining quenches from a holographic perspective using AdS/BCFT. We find that the tip of the EOW brane in the bulk (both in the vacuum or at finite temperature/size) moves radially according to the equation of motion of a massive particle in the Bañados geometry. If we associate to the tip of the EOW brane a particle with mass $m=c/16$, its proper radial momentum is proportional to the rate of growth of the spread complexity in the boundary CFT. Although this relation differs by a factor of $1/2$ from the result found in \cite{Caputa:2024sux} for a local operator quench and its holographic dual, it provides one of the first explicit examples where a well-defined measure of complexity can be directly matched with bulk momentum in $AdS_3$. 

The results reported in this paper open several interesting avenues for future research.
The local quench dynamics of a different complexity measure, the circuit complexity, has been studied in \cite{DiGiulio:2021noo} for systems of coupled harmonic oscillators. Exploiting Nielsen's approach \cite{Dowling:2006tnk} and its application in the context of bosonic Gaussian states \cite{Jefferson:2017sdb}, numerical results on the circuit complexity have shown the same time dependence as the spread complexity \eqref{eq:spreadcomplexity}, but with a different prefactor. Understanding the relations between these works would contribute to further mapping of possible connections between various types of complexity measures and how they probe the out-of-equilibrium quantum dynamics.

Our analysis focused on dynamics after local quenches in a single point of the CFT state. These protocols can be generalized to double (or multiple) local quenches, where joinings, splittings, and local operator insertions occur in two or more points of the initially prepared system. Interesting results on the entanglement evolution along such dynamics have been discussed in the literature \cite{Caputa:2019avh,Kusuki:2019avm}. Studying the double local quench dynamics using Krylov space tools provides an exciting future exploration, in particular concerning the role of gravitational forces in the holographically dual scenarios \cite{Caputa:2019avh} and how they can be detected using complexity.

Last but not least, we have analyzed the local quench dynamics from a Krylov space perspective. Other insightful protocols to study the out-of-equilibrium quantum dynamics are the global quantum quenches \cite{Calabrese:2005in}. Despite various works on characterizing global quench dynamics with Krylov space tools \cite{Pal:2023yik,Gautam:2023bcm,Camargo:2024rrj}, how the universal properties along critical evolutions manifest in the spread complexity and K-entropy is unclear.
Similarly to the local quenches, the global quench dynamics can be formulated in a CFT language \cite{Calabrese:2005in,Calabrese:2016xau,Das:2023xaw}. It would be interesting to perform a CFT analysis of Krylov space quantities and their evolution after a global quench. In particular, studying the quench evolution of the spread complexity starting from structured initial states, such as cross-cap states \cite{Chalas:2024yts,Wei:2024kkp}, is expected to bring insights into the out-of-equilibrium behaviour of quantum correlations.

\bigskip
\noindent {\bf \large Acknowledgments}
\\
We are grateful to Filiberto Ares, Jerome Dubail, Javier Magan, Alexey Milekhin, Sara Murciano, Joan Simon and Zixia Wei for discussions and comments on the draft. This work is supported by the ERC Consolidator grant (number: 101125449/acronym: QComplexity).  Views and opinions expressed are however those of the authors only and do not necessarily reflect those of the European Union or the European Research Council. Neither the European Union nor the granting authority can be held responsible for them. P. Caputa is supported by the NCN Sonata Bis 9 2019/34/E/ST2/00123 grant.

\appendix

\section{An alternative method for the return amplitude}
\label{app:CFTdetails}
In this appendix, we discuss an alternative method to derive the CFT return amplitude \eqref{eq:Snormalized}. This approach helps generalize \eqref{eq:Snormalized} to other interesting scenarios.

\subsection{Joining quench}
\label{sec:returnamplitudeCFT}

\begin{figure}[h!]
\centering
\includegraphics[width=1.\textwidth]{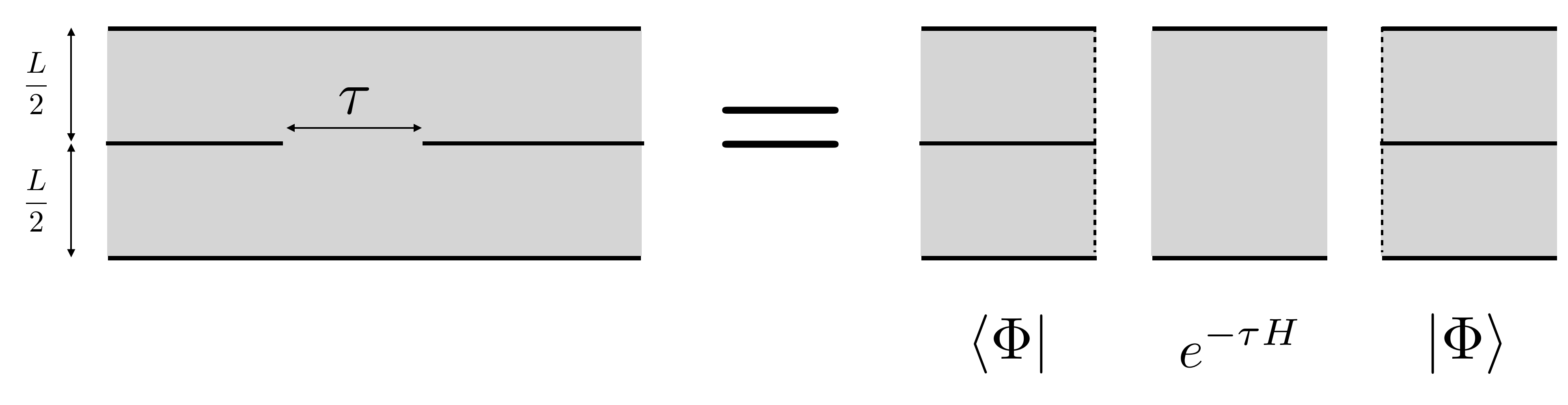}
\caption{The Euclidean return amplitude (\ref{eq:returnamplitudeEuclidean}) is computed as the CFT partition function on the geometry in the left panel given by a doubled infinite strip with a slit of size $\tau$. As depicted in the right panel, this geometry can be decomposed into three parts. In the spirit of radial quantization, the semi-infinite left and right parts give rise to the state $|\Phi\rangle$ and its complex conjugate, and the finite central part corresponds to the insertion of the evolution operator $e^{-\tau H}$.}
\label{fig:doublepants}
\end{figure}

\begin{figure}[t!]
\centering
\includegraphics[width=1.\textwidth]{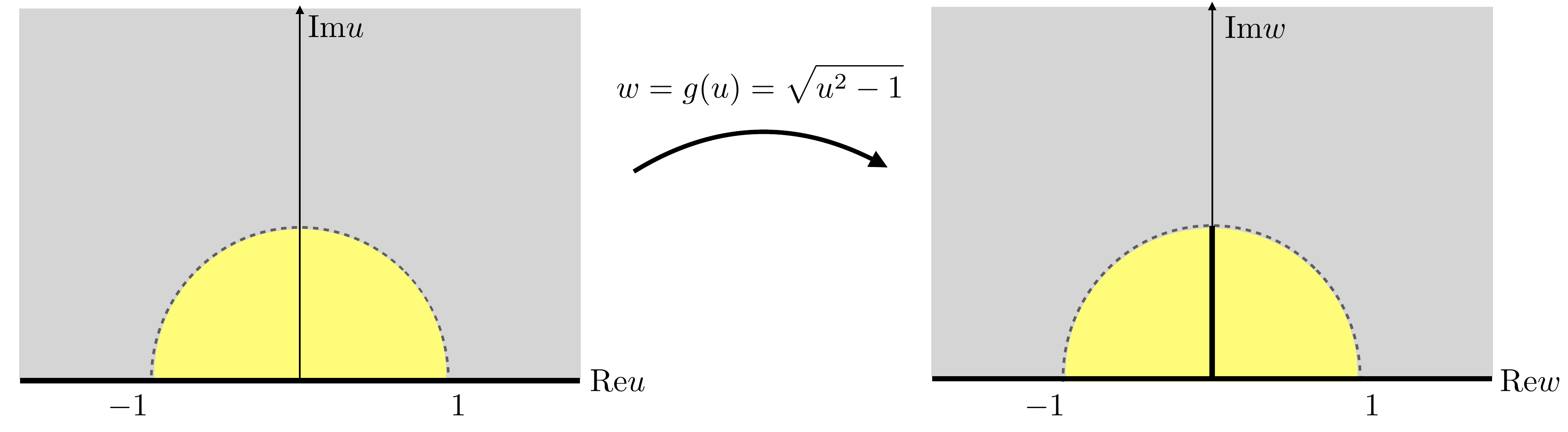}
\caption{The left and right geometries depicted in the right panel of Fig.\,\ref{fig:doublepants} are conformally equivalent to the half-plane with a vertical slit, represented in the right panel of this figure. This geometry can be further mapped by (\ref{eq:mapping}) to the half-plane without slits (left panel). The yellow regions describe the preparation of the vacuum state $|0\rangle$ (left panel) and $|\Phi\rangle$ (right panel).}
\label{fig:conformalmapping}
\end{figure}

Also for this method, the starting point to determine the return amplitude is \eqref{eq:returnamplitudeEuclidean}. The idea now is to think of the geometry in the left panel of Fig.\,\ref{fig:JoiningQuenchGeom} as decomposed into three parts, as pictorially represented in Fig.\,\ref{fig:doublepants}.
The left and the right parts represent the preparation of a state $|\Phi\rangle$  and its conjugate, while the central part is given by the Euclidean evolution induced by $e^{-\tau H}$. Thus, the function $Z(\tau)$ in \eqref{eq:returnamplitudeEuclidean} can be written as
\begin{equation}
\label{eq:returnamplitudeEuclidean_v2}
   Z(\tau)=  \langle \Phi|e^{-\tau H}|\Phi \rangle\,.
\end{equation}
Comparing this expression with (\ref{eq:returnamplitudeEuclidean}), we interpret the state $|\Phi\rangle$ as an explicit realization of $\lim_{\lambda\to\infty}e^{-\lambda(H_{L/2}\otimes \boldsymbol{1}_B+\boldsymbol{1}_A\otimes H_{L/2})}|s\rangle$. To find a useful expression for $|\Phi\rangle$, let us study the geometry obtained by preparing this state via path integral. This geometry is conformally equivalent to the half-plane with a vertical slit of height equal to one (right panel in Fig.\,\ref{fig:conformalmapping}).
Now we apply the main idea, which is to perform a conformal transformation $w=g(u)$, which deforms the half-plane with the vertical slit and coordinates $u$ into the half-plane geometry with coordinates $w$. This last geometry is obtained by preparing the CFT vacuum $|0\rangle$ through path integral.
In CFT, any conformal mapping corresponds to a combination of Virasoro generators. If the mapping relates the geometry where the vacuum state is prepared to the one where $|\Phi\rangle$ is prepared, by applying the aforementioned combination of Virasoro generators to $|0\rangle$ we obtain the state $|\Phi\rangle$.
As discussed in \cite{Dubail:2010arf}, the conformal mapping relating the two geometries in Fig.\,\ref{fig:conformalmapping} is given by
\begin{equation}
\label{eq:mapping}
   w=g(u)=\sqrt{u^2-1}\,,
\end{equation}
which leads to the state
\begin{equation}
\label{eq:trouserfromvacuum}
    |\Phi\rangle=e^{-\frac{1}{2}L_{-2}}|0\rangle\,,
\end{equation}
to plug in \eqref{eq:returnamplitudeEuclidean_v2}.
In the half-plane geometry, the evolution Hamiltonian can be written as
\begin{equation}
\label{eq:HamiltonianBCFT}
    H=
    \frac{1}{2\pi}\int_{0}^L \left( T(x,\tau)+\bar{T}(x, \tau)\right)dx
    =\frac{\pi}{L}\left(L_0-\frac{c}{24}\right)
    \,,
\end{equation}
where the absence of $\bar{L}_0$ in the last expression is due to the presence of the boundary.
Thus, we can write
\begin{equation}
\label{eq:returnamplitude_Virasoro}
    Z(\tau)=  e^{\frac{\pi c \tau}{24 L}}\langle 0|e^{-\frac{1}{2}L_{2}}e^{-\frac{\pi\tau}{L}L_0}e^{-\frac{1}{2}L_{-2}}|0\rangle\,.
\end{equation}
Expanding the left and the right exponentials,
\begin{equation}
    Z(\tau)= e^{\frac{\pi c \tau}{24 L}}\sum_{m,n=0}^\infty\frac{1}{m!n!}\left(\frac{-1}{2}\right)^{m+n}
    \langle 0|L_2^m e^{-\frac{\pi\tau}{L}L_0}L_{-2}^n|0\rangle\,.
\end{equation}
The expectation value in this expression can be evaluated by exploiting the Virasoro algebra. Indeed, since
\begin{equation}
  L_0 L_{-2}^n|0\rangle=2n L_{-2}^n|0\rangle \,,
\end{equation}
 for any function $f$ admitting a Taylor expansion, we can write
\begin{equation}
   f(L_0) L_{-2}^n|0\rangle=f(2n) L_{-2}^n|0\rangle  \,.
\end{equation}
Thus, we have
\begin{equation}
   Z(\tau)= e^{\frac{\pi c \tau}{24 L}}\sum_{m,n=0}^\infty\frac{1}{m!n!}\left(-\frac{1}{2}\right)^{m+n} e^{-\frac{2\pi\tau}{L}n}
    \langle 0|L_2^m L_{-2}^n|0\rangle \,.
\end{equation}
Using that \cite{Dubail:2010arf}
\begin{equation}
   \langle 0|L_2^m L_{-2}^n|0\rangle=\delta_{mn}\frac{n!}{2^n}\prod_{p=0}^{n-1} (8p+c)\,,
\end{equation}
we finally find
\begin{equation}
\label{eq:Stau_Euclidean_final}
   Z(\tau)= e^{\frac{\pi c \tau}{24 L}}
   \sum_{n=0}^\infty\frac{1}{n! 8^n} e^{-\frac{2\pi\tau}{L}n} \prod_{p=0}^{n-1} (8p+c)
   =
   e^{\frac{\pi c \tau}{6 L}}\left(2\sinh\frac{\pi \tau}{L}\right)^{-c/8}
    \,.
\end{equation}
Notice that the series in the middle expression 
can be resummed only if we keep $e^{\frac{2\pi\tau}{L}n}$ and not if we replace $e^{\frac{2\pi\tau}{L}n}$ with $\left(\frac{2\pi\tau}{L}n\right)^k$, $k>0$ originated by its Taylor expansion. This is related to the fact that \eqref{eq:Stau_Euclidean_final} diverges when $\tau\to 0$. Performing the Wick's rotation
$\tau=-{\rm i}t+2\varepsilon$, and dividing the result by its value at $t=0$, we reproduce exactly
\eqref{eq:Snormalized}.

\subsection{Other local quenches in CFT}
\label{subapp:otherquenches}
As anticipated, the approach presented in this appendix allows us to generalize the evolution \eqref{eq:Snormalized} of the return amplitude to other instances of local joining quenches. In what follows, we consider two examples of these protocols.

\begin{enumerate}
    
    \item At time $t=0$ the system is in a CFT ground state on an open interval of length $L$. For $t>0$, we let the system evolve with the Hamiltonian of a CFT on a cylinder with compact spatial dimension of length $L$. This protocol can be seen as a quench from an open to a periodic system. 

\item At time $t=0$ we have two CFTs at finite temperature $\beta^{-1}$ on half-infinite lines. For times $t>0$, we let the finite temperature system evolve with a Hamiltonian defined on the full line, i.e. at time $t=0$ we join the two half-lines.

\end{enumerate}

\begin{figure}[t!]
\centering
\includegraphics[width=1.\textwidth]{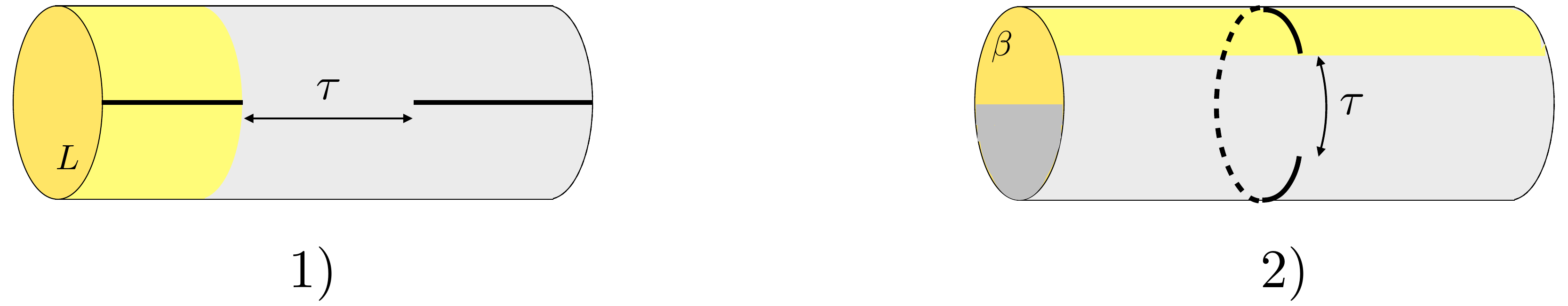}
\caption{The CFT partition functions on the two-dimensional geometries in this figure allow us to compute the return amplitudes after the two local quenches listed in this section. The enumeration in this figure matches the one in the text for a simple identification. The two cylinders are infinite, and the circumferences along the compact directions are reported on each panel. The thick black lines are defects introduced in the geometry by the initial non-translational invariant state. The yellow sub-geometries are the geometrizations of the states $|\Phi_i\rangle$ in (\ref{eq:generalPartitionFunction}).}
\label{fig:3otherGeometries}
\end{figure}

As explained in Sec.\,\ref{sec:returnamplitudeCFT}, the return amplitudes in these three cases can be computed as CFT partition functions in specific geometries. These geometries are shown in Fig.\,\ref{fig:3otherGeometries}, where the enumeration corresponds to the one of the list above. Notice that all the cylinders in Fig.\,\ref{fig:3otherGeometries} are infinite.

Paralleling the computation in Sec.\,\ref{sec:returnamplitudeCFT}, we want to write the partition functions on the geometries in Fig.\,\ref{fig:3otherGeometries} as 
\begin{equation}
\label{eq:generalPartitionFunction}
    Z^{(j)}(\tau)=\langle\Phi_j|e^{-\tau H^{(j)}}|\Phi_j\rangle\,,
    \qquad
    j=1,\,2\,,
\end{equation}
where $|\Phi_j\rangle$ is the state whose geometrization is given by the yellow sub-geometry in the panel $j)$ in Fig.\,\ref{fig:3otherGeometries}, and $H^{(j)}$ is the CFT evolution Hamiltonian on such sub-geometry. Once the partition function is known, we obtain the return amplitude in real time as
\begin{equation}
\label{eq:Sgeneral}
   S^{(j)}(t)=\bigg|\frac{Z^{(j)}(-{\rm i}t +2\varepsilon)}{Z^{(j)}(2\varepsilon)}\bigg| \,,
\end{equation}
where, as discussed in Sec.\,\ref{subsec:returnampl_CFT}, we can take the absolute value, i.e. get rid of the phases, by shifting the ground state energy of the evolution Hamiltonian.
In the following subsections, we work out the partition function (\ref{eq:generalPartitionFunction}) in the two cases of interest. 

\subsubsection{Local quench from open to periodic systems}
\label{subsec:fromopentoperiodic}
\begin{figure}[t!]
\centering
\includegraphics[width=1.\textwidth]{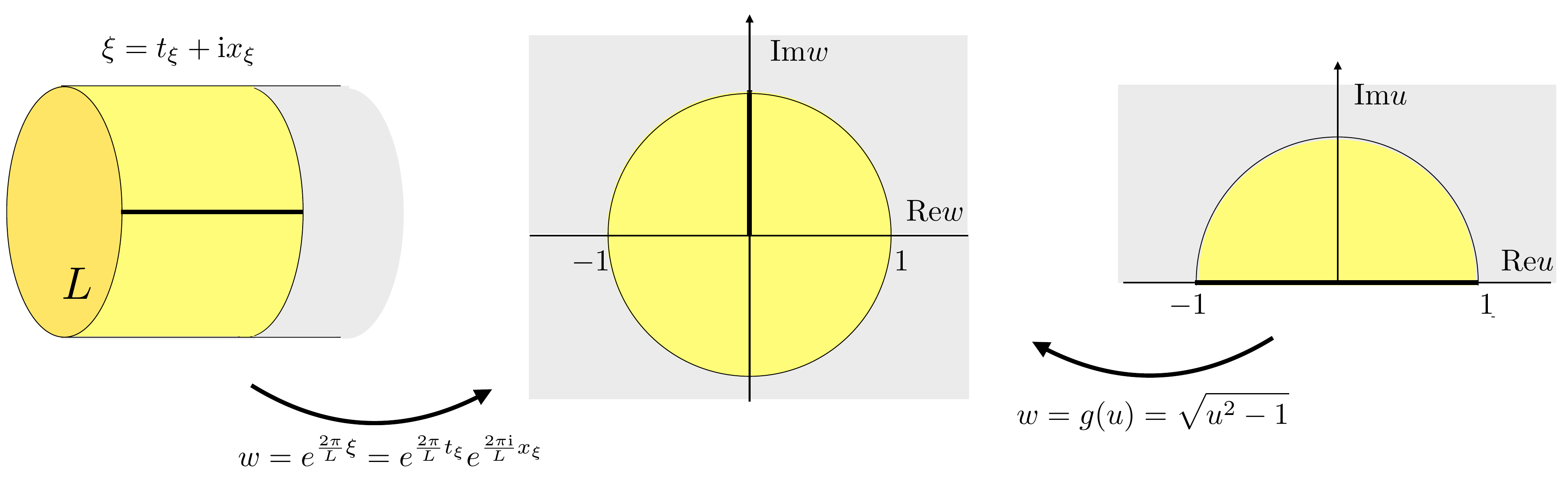}
\caption{The yellow cylinder in the left panel is the yellow sub-geometry in the left panel of Fig.\,\ref{fig:3otherGeometries}. We map it first to the unit disk with a defect along the upper imaginary axis (middle panel) and finally to the half disk with a defect along the diameter (right panel). This last geometry is the same depicted in the left panel of Fig.\,\ref{fig:conformalmapping} and, therefore, the return amplitude after this local quench protocol is given by (\ref{eq:Snormalized}).}
\label{fig:QuenchFromOpentoPeriodic}
\end{figure}

We begin from the local quench protocol 1) in the list above. 
The state $|\Phi_1\rangle$ in (\ref{eq:generalPartitionFunction}) can be represented geometrically as the yellow sub-geometry in the left panel
of Fig.\,\ref{fig:3otherGeometries}, also reported in the left panel of Fig.\,\ref{fig:QuenchFromOpentoPeriodic}. If we parametrize this half-cylinder geometry with coordinates $\xi=t_\xi+{\rm i} x_\xi$, with $t_\xi\in \mathbb{R}$ and $x_\xi\in [0,L)$, we can map it to the unit disk with a defect along the upper part of the imaginary axis (see the middle panel in Fig.\,\ref{fig:QuenchFromOpentoPeriodic}) through the transformation
\begin{equation}
\label{eq:mappingRadialQuant}
   w=e^{\frac{2\pi}{L}\xi}=e^{\frac{2\pi}{L}t_\xi}e^{\frac{2\pi{\rm i}}{L}x_\xi}\,.
\end{equation}
Finally, applying $g^{-1}$, where $g$ is defined in (\ref{eq:mapping}), we obtain a half disk with the defect mapped along the diameter.
This is the state geometric representation that we have used to describe $|\Phi\rangle$ in Sec.\,\ref{sec:returnamplitudeCFT} (see the left panel of Fig.\,\ref{fig:conformalmapping}). Thus, the Hamiltonian $H^{(1)}$ is represented as (\ref{eq:HamiltonianBCFT}), the partition function in Euclidean time is given by (\ref{eq:Stau_Euclidean_final}) and the return amplitude is (\ref{eq:Snormalized}), up to a redefinition of the non-universal cutoff.

\subsubsection{Joining quench of an infinite system at finite temperature}
For the local quench obtained joining two half-systems described by a CFT at finite temperature $\beta^{-1}$, it is enough to notice that the geometry 2) in Fig.\,\ref{fig:3otherGeometries} can be mapped to the geometry 1). To see this, let us we call $\zeta={\rm i}t_\zeta+x_\zeta$, with $x_\zeta\in \mathbb{R}$ and $t_\zeta\in [0,\beta)$, the coordinates of the geometry 2). We first consider the mapping to the complex plane
\begin{equation}
   \zeta\to e^{\frac{2\pi}{\beta}\zeta} \,,
\end{equation}
then we substitute
 $t_\zeta\to t_\xi $, $x_\zeta\to x_\xi $ and  $\beta\to{-\rm i} L$, and finally we map it back to the cylinder obtaining the geometry 1) in Fig.\,\ref{fig:3otherGeometries} with coordinates $\xi=t_\xi+{\rm i} x_\xi$, with $t_\xi\in \mathbb{R}$ and $x_\xi\in [0,L)$. Notice that, in this new geometry, the width $\tau$ of the slit is unchanged. Thus, the return amplitude after this process is obtained by replacing $L\to{\rm i} \beta$ in (\ref{eq:Snormalized}), which leads to (\ref{eq:SfiniteT}).

\section{Return amplitudes after local quenches in free chains}
\label{app:lattice}
In this appendix, we discuss how to compute the return amplitudes after local quenches in fermionic and bosonic free chains. These techniques have been exploited in the main text to obtain the data in Figs.\,\ref{fig:localquenchFreeChains} and \ref{fig:SplittingReturnAmpl}, to benchmark the CFT results reported in the manuscript.

\subsection{Free fermionic chains}
\label{subapx:FF}

\subsubsection{Joining quench in free fermionic chains: ground state}
Consider a chain of free fermions made by $L$ sites with Hamiltonian
\begin{equation}
\label{eq:FFHamiltonian}
  H_\alpha\equiv-\frac{1}{2}\sum_{j=1}^L \alpha_j\left( c_j^\dagger c_{j+1}+ c_{j+1}^\dagger c_{j}\right)  \,,
\end{equation}
where $\{c_j^\dagger,c_i\}=\delta_{ij}$, $\{c_j,c_i\}=\{c_j^\dagger,c^\dagger_i\}=0$ and $\alpha_{L/2}=\alpha$, $\alpha_{j\neq L/2}=1$. 
We impose open boundary conditions at the beginning and end of the system.
The local joining quench is realized by preparing the system at time $t=0$ in the ground state of $H_0$, i.e. two decoupled chains with $L/2$ sites and open boundary conditions. If we denote by $|{\rm GS}_L\rangle$ the ground state of a homogeneous chain with $L$ sites and open boundary conditions, the initial state we choose reads
\begin{equation}
\label{eq:initial state}
  |\psi(0)\rangle= |{\rm GS}_{L/2}\rangle\otimes|{\rm GS}_{L/2}\rangle \,.
\end{equation}
For $t>0$, we let the system evolve with $H_1$, i.e. we suddenly couple the two fermionic chains, inducing the evolution on the $L$-sites chain.
The initial state \eqref{eq:initial state} is a Gaussian state, and the evolution Hamiltonian is quadratic in the fermionic operators. Thus, the full dynamics is encoded in the time evolution of the two-point correlation functions. 

We start from the two-point functions in the ground state of a chain with $L$ sites. To compute them, we diagonalize the Hamiltonian $H_1$, which is equivalent, modulo changing the number of sites, to diagonalize the two half chains into which the system is initially divided.
Using the {\it sine} Fourier transform due to the lack of translational invariance,
\begin{equation}
\label{eq:FT}
    c_j=\sqrt{\frac{2}{L+1}}\sum_{q=1}^{L}\sin\left(\frac{\pi q j}{L+1} \right)c_q\,,
\end{equation}
the Hamiltonian acquires the diagonal form
\begin{equation}
\label{eq:diagonalHam}
    H_1=-\sum_{q=1}^{L}\cos\left(\frac{\pi q j}{L+1} \right)c_q^\dagger c_q\,.
\end{equation}
From this form, we construct the ground state $|{\rm GS}_{L}\rangle$ such that $c_q|{\rm GS}_{L}\rangle=0$ for any $q=1,\dots,L$.
For convenience, we collect all the two-point functions in the correlation matrix
\begin{equation}
 C^{(L)}_{ij}\equiv\langle {\rm GS}_{L}| c_i^\dagger c_j |{\rm GS}_{L}\rangle\,,
\end{equation}
$i,j=1,\dots, L$.
Using (\ref{eq:FT}) and the properties of the ground state, we obtain \cite{Alba_2014}
\begin{equation}
\label{eq:2ptsfunction_FF_Lchain}
 C^{(L)}_{ij}=\frac{1}{2(L+1)}\left[\frac{\sin\left(\frac{\pi(i-j)}{2}\right)}{\sin\left(\frac{\pi (i-j)}{2(L+1)}\right)}-\frac{\sin\left(\frac{\pi(i+j)}{2}\right)}{\sin\left(\frac{\pi (i+j)}{2(L+1)}\right)}\right]\,,   
\end{equation}
where, due to the lack of translational invariance, the two-point functions do not depend on the different $i-j$ only.

\subsubsection{Joining quench in free fermionic chains: time evolution}
\label{subsec:timeevolutionjoin}
The initial state (\ref{eq:initial state}) has an $L\times L$ correlation matrix with the following block structure
\begin{equation}
    C(0)=\begin{pmatrix}
          C^{(L/2)} & \boldsymbol{0}
        \\
        \boldsymbol{0}  & C^{(L/2)}
    \end{pmatrix}\,,
\label{eq:initial correlation}
\end{equation}
where $\boldsymbol{0}$ is the $L/2\times L/2$ matrix with zeroes in all the entries.
The correlation matrix at any time of the evolution induced by $H_1$ is
\begin{equation}
\label{eq:time_correlation}
  C_{ij}(t)=\langle \psi(t)| c_i^\dagger c_j |\psi(t)\rangle  \,,
\end{equation}
where
\begin{equation}
  |\psi(t)\rangle= e^{-{\rm i}H_1 t} |\psi(0)\rangle \,.
\end{equation}
Notice that (\ref{eq:time_correlation}) reduces to (\ref{eq:initial correlation}) when $t=0$.
We can rewrite $C(t)$ in a different useful way
\begin{equation}
\label{eq:time_correlation_v2}
 C_{ij}(t)= \langle \psi(0)| c_i^\dagger (t)c_j (t)|\psi(0)\rangle  \,,
\end{equation}
where
\begin{equation}
    c_i (t)= e^{{\rm i}H_1 t}c_i e^{-{\rm i}H_1 t}\,.
\end{equation}
To derive its explicit expression, we observe that since
\begin{equation}
    [c_q,H_1]=-\cos\left(\frac{\pi q}{L+1}\right)c_q\,,
\end{equation}
we easily find
\begin{equation}
    c_q (t)= e^{{\rm i}H_1 t}c_q e^{-{\rm i}H_1 t}=
    e^{{\rm i}t\cos\left(\frac{\pi q}{L+1}\right)} c_q\,.
\end{equation}
Performing the inverse Fourier transform, we obtain
\begin{equation}
\label{eq:timeevolutionfermion}
    c_i(t)=\frac{2}{L+1}\sum_{q,m=1}^L
    \sin\left(\frac{\pi q i}{L+1}\right)
    \sin\left(\frac{\pi q m}{L+1}\right)e^{{\rm i}t\cos\left(\frac{\pi q}{L+1}\right)}c_m
    =\sum_{m=1}^L U_{im}c_m\,,
\end{equation}
where we have introduced the $L\times L$ time evolution matrix
\begin{equation}
\label{eq:evolutionUmat}
    [U(t)]_{im}\equiv \frac{2}{L+1}\sum_{q=1}^L
    \sin\left(\frac{\pi q i}{L+1}\right)
    \sin\left(\frac{\pi q m}{L+1}\right)e^{{\rm i}t\cos\left(\frac{\pi q}{L+1}\right)}\,.
\end{equation}
Plugging (\ref{eq:timeevolutionfermion}) into (\ref{eq:time_correlation_v2}), we can rewrite the time-evolved correlation matrix in the following convenient way \cite{Eisler_2007}
\begin{equation}
\label{eq:corr_mat_evolved}
    C(t)=U^\dagger(t)C(0) U(t)\,,
\end{equation}
where the initial correlation matrix $C(0)$ is defined in (\ref{eq:initial correlation}).

\subsubsection{Splitting quench in free fermionic chains}
\label{subapp:SplittingFF}
The splitting quench in this setup consists of starting from an initial state given by the ground state of $H_1$ in \eqref{eq:FFHamiltonian}, i.e.
\begin{equation}
\label{eq:initial state_split_FF}
  |\psi(0)\rangle= |{\rm GS}_{L}\rangle \,,
\end{equation}
 and letting the system evolve with the Hamiltonian $H_0$ in \eqref{eq:FFHamiltonian} describing two disjoint chains with $L/2$ sites each. As already commented on for the joining quench, this splitting quench evolution involves Gaussian states at any time $t$. Thus, it can be fully characterized by the evolution of the two-point correlation functions of the fermionic degrees of freedom.
 The correlation matrix \eqref{eq:time_correlation} of the initial state is simply given by 
 \begin{equation}
    C(0)=C^{(L)} \,,
 \end{equation}
 where the entries of $C^{(L)}$ are given by \eqref{eq:2ptsfunction_FF_Lchain}.
 Since the time evolution is given by $H_0$, the corresponding matrix inducing the evolution on the initial correlation matrix has a block diagonal structure. This is consistent with the fact that $H_0$ does not couple the left and right parts of the chain. Adapting the computation in Appendix \,\ref{subsec:timeevolutionjoin}, we find that $C(t)$ is given by \eqref{eq:corr_mat_evolved}, where now the matrix $U(t)$ is given by
 \begin{equation}
 \label{eq:evolutionUmat_split}
    U(t)=
    \begin{pmatrix}
          \tilde{U}(t) & \boldsymbol{0}
        \\
        \boldsymbol{0}  & \tilde{U}(t)
    \end{pmatrix} \,,
 \end{equation}
 and $\tilde{U}(t)$ is given by \eqref{eq:evolutionUmat} with the replacement $L\to L/2$.
 
\subsubsection{Return amplitudes in free fermionic systems}
The absolute value of the return amplitude can be seen as the fidelity between the initial state and the state evolved at time $t$.
The fidelity between two fermionic pure Gaussian states $|\psi\rangle, |\phi\rangle$ with the only non-vanishing two-point correlations being of the form $ \langle c_i^\dagger c_j\rangle$ has the simple expression \cite{Parez:2022sgc}
\begin{equation}
   |\langle\psi|\phi\rangle|=
   \sqrt{\det\left(\frac{\boldsymbol{1}+J_\psi J_\phi}{2}\right)}\,,
\end{equation}
where
\begin{equation}
    J_\psi \equiv 2 C_\psi-\boldsymbol{1}\,,
    \qquad
    J_\phi \equiv 2 C_\phi-\boldsymbol{1}\,,
\end{equation}
and
\begin{equation}
    (C_\psi)_{ij} \equiv \langle\psi| c_i^\dagger c_j |\psi\rangle\,,
    \qquad
    (C_\phi)_{ij} \equiv \langle\phi| c_i^\dagger c_j |\phi\rangle\,,
\end{equation}
contain the (only non-vanishing) correlation functions in the two considered states.
This formula can be applied to the initial state, and the state evolved at time $t$ after the local quench, given that these states are both Gaussian. The result is 
\begin{equation}
\label{eq:return ampli_FF}
    |S(t)|= |\langle\psi(t)|\psi (0)\rangle|
    =
\sqrt{\det\left(\frac{\boldsymbol{1}+J(t) J(0)}{2}\right)}\,,
\end{equation}
with
\begin{equation}
\label{eq:Jmat_FF}
     J(t) \equiv 2 C(t)-\boldsymbol{1}\,.
\end{equation}
The matrix $C(t)$ in \eqref{eq:Jmat_FF} is obtained from (\ref{eq:corr_mat_evolved}) with $U(t)$ in \eqref{eq:evolutionUmat} if we consider a joining quench and in \eqref{eq:evolutionUmat_split} for a splitting quench.
As commented in the main text, possible phases in the expression of the return amplitude can be removed by an appropriate shift of the ground state energy of the system undergoing the quench evolution. Thus, we focus only on the absolute value of the return amplitude, exploiting the formula \eqref{eq:return ampli_FF} for checking the CFT predictions.
Indeed, \eqref{eq:return ampli_FF} has been used to obtain the data points in the left panel of Fig.\,\ref{fig:localquenchFreeChains} and in Fig.\,\ref{fig:SplittingReturnAmpl}.

\subsection{Harmonic chain}
\label{subapx:HC}

\subsubsection{Joining quench in harmonic chains: ground state}
Consider the Hamiltonian of a harmonic chain with $L$ sites 
\begin{equation}
  \label{eq:HC Hamiltonian}
    H_L(\omega)=\frac{1}{2}\sum_{j=0}^{L}\left[
    p_j^2+\omega^2q_j^2+(q_{j+1}-q_j)^2
    \right]\,,
\end{equation}
where the operators $p_j$ and $q_j$ satisfies the canonical commutation relations
\begin{equation}
[q_i,p_j]={\rm i}\delta_{ij}\,,
\qquad
[q_i,q_j]=[p_i,p_j]=0\,,
\end{equation}
and $\omega$ is the frequency parameter, which becomes the mass in the continuum limit. Thus, the model becomes gap-less (and conformal in the continuum limit) when $\omega\to 0$.
We impose Dirichlet boundary conditions (DBC), meaning that $p_0=p_{L+1}=0$ and $q_0=q_{L+1}=0$.
The Hamiltonian (\ref{eq:HC Hamiltonian}) is diagonalized by introducing the creation and annihilation operators $a_k$ and $a_k^\dagger$. We find
\begin{equation}
H_L(\omega)=\sum_{k=1}^L   \Omega_k \left(a_{k}^\dagger a_{k}-\frac{1}{2}\right)\,,    
\end{equation}
where 
\begin{equation}
    \Omega_k=\sqrt{\omega^2+4 \left[\sin\left(\frac{\pi k}{2(L+1)}\right)\right]^2},
\end{equation}
provides the dispersion relation.
The ground state $|0_L\rangle$ of this model is Gaussian, and, therefore, it is fully described by its two-point correlation functions, collected into the covariance matrix. Given a Gaussian state $|\phi\rangle$, its $2L\times 2L$ covariance matrix  $\gamma$ has entries
\begin{equation}
    \gamma_{ij}\equiv {\rm Re}\left[\langle \phi|\boldsymbol{r}_i \boldsymbol{r}_j |\phi\rangle\right]\,,
\end{equation}
where we have collected all the operators $p_j$ and $q_j$ into the vector $\boldsymbol{r}\equiv (q_1,\dots, q_L, p_1,\dots, p_L)^{\textrm{t}}$.

\subsubsection{Joining quench in harmonic chains: time evolution}
The joining quench in this model is realized by considering the initial state to be 
\begin{equation}
  |\psi(0)\rangle= |0_{L/2}\rangle\otimes |0_{L/2}\rangle \,,
\end{equation}
namely the product of the ground states of two decoupled half-chains (\ref{eq:HC Hamiltonian}) of length $L/2$ (we maintain DBC in all the endpoints).
For $t>0$, we let the system evolve with $H_L(\omega)$, i.e. we join together the two initial parts, namely
\begin{equation}
\label{eq:initialstate_boson}
  |\psi(t)\rangle= e^{-{\rm i}H_L(\omega)t}|\psi(0)\rangle \,.
\end{equation}
Given that the initial state is Gaussian and the evolution Hamiltonian is quadratic, $|\psi(t)\rangle$ is also Gaussian, and its
evolution is fully described by the evolution of the two-point correlation functions.
The covariance matrix evolved at time $t$ can be written in terms of the initial one as \cite{Eisler:2014dze}
\begin{equation}
\label{eq:covariancetime_t}
    \gamma(t)= E(t) \gamma(0) E(t)^{\rm t}\,,
\end{equation}
where $E(t)$ is given by
\begin{equation}
\label{eq:evolution boson}
    E(t)=\left(V_L^{\rm t}\oplus V_L^{\rm t} \right) \mathcal{E}(t)\left(V_L \oplus V_L\right)\,,
    \qquad
    \mathcal{E}(t)=\begin{pmatrix}
       \mathcal{D}(t)
        &
        \mathcal{A}(t)
        \\
        \mathcal{B}(t)
        &
        \mathcal{D}(t)
    \end{pmatrix}\,,
\end{equation}
with $\mathcal{A}$, $\mathcal{B}$ and $\mathcal{D}$ diagonal matrices with $L$ entries
\begin{eqnarray}
\mathcal{A}_k(t)&=&\frac{\sin(\Omega_k t)}{\Omega_k}\,,
\\
\mathcal{B}_k(t)&=&-\Omega_k\sin(\Omega_k t)\,,
\\
\mathcal{D}_k(t)&=&\cos(\Omega_k t)\,,
\end{eqnarray}
and
\begin{equation}
\label{eq:FTmatrix}
    (V_L)_{jk}=\sqrt{\frac{2}{L+1}}\sin\left(\frac{\pi jk}{L+1}\right)\,.
\end{equation}
Finally, we need the covariance matrix of the initial state (\ref{eq:initialstate_boson}). It reads \cite{Eisler:2014dze}
\begin{equation}
\label{eq:covariance_initial}
    \gamma(0)=\frac{1}{2}\left(V^{\rm t}_{L/2}\oplus V^{\rm t}_{L/2}\oplus V^{\rm t}_{L/2}\oplus V^{\rm t}_{L/2}\right)\left(T^{-1}\oplus T\right)\left(V_{L/2}\oplus V_{L/2}\oplus V_{L/2}\oplus V_{L/2}\right)\,,
\end{equation}
where $T$ is the $L\times L$ diagonal matrix
\begin{equation}
   T={\rm diag}\left(\Omega_1,\dots\Omega_{L/2},\Omega_1,\dots\Omega_{L/2}\right)\,,
\end{equation}
and $V_{L/2}$ is given in (\ref{eq:FTmatrix}).
Plugging (\ref{eq:evolution boson}) and (\ref{eq:covariance_initial}) into (\ref{eq:covariancetime_t}), we obtain the evolution of the covariance matrix along the joining quench.

\subsubsection{Return amplitudes in bosonic Gaussian states}
As already seen for the fermionic chains, also for this bosonic model, we compute the absolute value of the return amplitude as the fidelity between the initial state $|\psi(0)\rangle$ and the evolved state $|\psi(t)\rangle$. Since the two states are Gaussian, the fidelity can be written in terms of their covariance matrices $\gamma(0)$ and $\gamma(t)$. It reads \cite{Banchi:2015rmr}
\begin{equation}
\label{eq:returnampl_boson}
|S(t)|=|\langle\psi(t)|\psi(0)\rangle|=\left[\det\left(\gamma(0)+\gamma(t)\right)\right]^{-1/4}\,.
\end{equation}
The absolute value of the return amplitude after the joining quench is evaluated by plugging (\ref{eq:covariancetime_t}) and (\ref{eq:covariance_initial}) into \eqref{eq:returnampl_boson}. Since we are interested in the comparison with the CFT predictions, we choose the $\omega=0$, which, in the continuum limit, corresponds to the evolution of a CFT with $c=1$. The formula \eqref{eq:returnampl_boson} has been used to obtain the data points reported in the right panel of Fig.\,\ref{fig:localquenchFreeChains}.

\section{Spread complexity after a domain wall quench}
\label{app:domainwallquench}
In this appendix, we compute the spread complexity and the K-entropy after a different type of local quench, the domain wall quench. This quench differs from the others discussed in this manuscript for a different form of the initial state. The evolution of the spread complexity found for this quench implies that the dynamics on the Krylov chain strongly depends on the choice of the initial state.

\subsection{Domain wall quench in free fermionic chains}
\begin{figure}[t!]
\centering
\includegraphics[width=1.\textwidth]{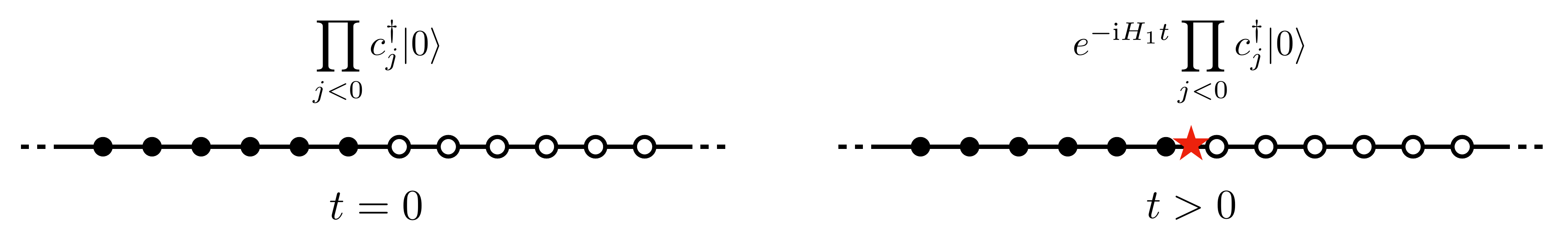}
\caption{Pictorial representation of the domain wall quench protocol. The initial state \eqref{eq:DWinitialstate} (left panel) is obtained by creating excitations on the left part of the vacuum state of a fermionic chain. We let this initial state evolve via the Hamiltonian \eqref{eq:diagonalHam} (right panel).
}
\label{fig:DWquenchsetup}
\end{figure}

Consider two fermionic chains, each with $L/2$ sites. We assume that the right half is initially in the vacuum, while, in the left part, we create an excitation on each of the $L/2$ sites. 
The initial state is, therefore, written as
\begin{equation}
\label{eq:DWinitialstate}
   |\psi(0)\rangle=\prod_{j<0}c_j^\dagger\vert 0\rangle \,.
\end{equation}
The state \eqref{eq:DWinitialstate} is called domain wall state.
At time $t>0$, we join the two chains and let the system evolve
via the Hamiltonian $H_1$ \eqref{eq:diagonalHam}. Given the peculiar initial state, this type of local quench is called domain wall quench. The protocol is pictorially represented in Fig.\,\ref{fig:DWquenchsetup}.

 We want to compute the return amplitude after the domain wall quench, which reads
\begin{eqnarray}
\label{eq:def_return amplitude}
 S_{\rm DW}(t)&=&\langle \psi(0) |e^{{\rm i}t H_1}|\psi(0)\rangle= \langle 0| c_{L/2}\dots c_{1}  e^{{\rm i}t H_1} c_{1}^\dagger\dots c_{L/2}^\dagger\vert 0\rangle 
 \\
 &=&
 \langle 0| c_{L/2} \dots c_{1}   c_{1}^\dagger(t)\dots c_{L/2}^\dagger(t)\vert 0\rangle
 \\
 &=&
 \det_{i,j\leq L/2} \langle 0|  c_{i}  c_{j}^\dagger(t)\vert 0\rangle
 \,,
\end{eqnarray}
where, in the last step, we have used Wick's theorem. Exploiting \eqref{eq:timeevolutionfermion} and the fermionic anti-commutation relations, we find
\begin{equation}
    \label{eq:return amplitude_det}
 S_{\rm DW}(t)=\det( U(t)^*)\,,
\end{equation}
where the evolution matrix $U(t)$ is defined in \eqref{eq:evolutionUmat}.

An elegant exact result can be found in the thermodynamic limit, i.e. $L\to\infty$ \cite{Viti:2015khg}. In this limit,
the entries of $U(t)$ become
\begin{equation}
    \label{eq:U mat_td limit}
 U_{ij}(t)=U_{i-j}(t)-U_{i+j}(t)\,,
\end{equation}
where
\begin{equation}
    \label{eq:U x}
 U_{x}(t)=\int_{-\pi}^\pi \frac{dq}{2\pi}e^{- {\rm i}q x }e^{{\rm i}t\cos\left(q\right)}\,.
\end{equation}
Crucially, the matrix $U(t)$ is written as the sum of a Toeplitz matrix and a Hankel matrix. Many expansions for Toeplitz and Hankel determinants are known in the limit where the matrix size has grown to infinity. Using some of these results, we find \cite{Viti:2015khg}
\begin{equation}
    \label{eq:S_tdlimit}
\lim_{L\to\infty}S_{\rm DW}(t)=e^{-\frac{t^2}{8}}\,.
\end{equation}
It is worth stressing that this expression is an exact lattice result and, therefore, is valid for any time $t$.

\subsection{Spread complexity and K-entropy}
In \cite{Balasubramanian:2022tpr}, it has been found that time evolution with Gaussian return amplitudes can be effectively described on the Krylov chain by a Hamiltonian with an emergent symmetry generated by a Heisenberg-Weyl algebra. We refer the interested reader to \cite{Balasubramanian:2022tpr} for the details. 
Given the form  \eqref{eq:S_tdlimit} of the return amplitude, we deduce that the domain wall local quench dynamics admits this effective description.
For these types of evolutions, the Lanczos algorithm has already been worked out explicitly, and the Lanczos coefficients that can be obtained from \eqref{eq:S_tdlimit} read
\begin{equation}
    \label{eq:LanczosDWquench}
a_n=0\,,\qquad b_n=\frac{\sqrt{n}}{2}\,.
\end{equation}
The corresponding spread complexity has the following quadratic time dependence
\begin{equation}
    \label{eq:spreadComplexity_DW}
C_\mathcal{K}(t)=\frac{t^2}{4}\,.
\end{equation}
Remarkably, the expression \eqref{eq:S_tdlimit} and the consequent spread complexity are exact results valid at any time scale, differently from the CFT predictions in Sec.\,\eqref{subsec:spreadComplexity} which are valid after a time $\varepsilon$. We find it, however, worth comparing the \eqref{eq:spreadComplexity_DW} with the result \eqref{eq:spreadcomplexity} found for the single joining quench of two semi-infinite CFT ground states. Indeed, the dependence on time is quadratic in both cases, signaling a similar spread of information after the two types of quench. In the CFT case, the prefactor of the quadratic growth is explicitly dependent on the cutoff $\varepsilon$, as opposed to \eqref{eq:spreadComplexity_DW}, which is valid at any time scale.

Leveraging the results in \cite{Balasubramanian:2022tpr}, we can analyze the K-entropy, which, for this evolution, reads 
\begin{equation}
    \label{eq:KentropyDW}
H_K(t)=\frac{t^2}{4}-\frac{t^2}{4}\ln\left(\frac{t^2}{4}\right)+\sum_{n=0}^\infty \frac{e^{-\frac{t^2}{4}}}{n!}\left(\frac{t^2}{4}\right)^n\,.
\end{equation}
The third term in \eqref{eq:KentropyDW} cannot be resumed, but we can study the early time regime, where the first two terms dominate, finding
\begin{equation}
    \label{eq:KentropyDW_early}
H_K(t)\simeq\frac{t^2}{4}-\frac{t^2}{4}\ln\left(\frac{t^2}{4}\right)\,.
\end{equation}
The K-entropy at time $t=0$ is vanishing as expected and has a quadratic initial growth.

\subsection{Domain wall quench with long range Hamiltonian}
Finally, we can consider the case of a domain wall local quench with evolution induced by the following Hamiltonian with long-range hoppings
\begin{equation}
    \label{eq:longrangeHamiltonian}
    H_{p}=\sum_{j=1}^L\sum_{l=1}^p u_l c^\dagger_{j+l}c_j+{\rm h.c.} \,.
\end{equation}
Diagonalizing \eqref{eq:longrangeHamiltonian} in the usual way, we find
\begin{equation}
    \label{eq:longrangeHamiltonian_diag}
    H_{p}=\sum_{k=1}^L \left[\sum_{l=1}^p 2u_l \cos\left(\frac{\pi k l}{L+1}\right)\right]  c^\dagger_{k}c_k \,.
\end{equation}
Considering the same initial state \eqref{eq:DWinitialstate} as before, the return amplitude is again given by \eqref{eq:return amplitude_det}, and the matrix $U(t)$ is the sum of a Toeplitz and a Hankel matrix. By adapting the computation above, it has been found that \cite{Viti:2015khg}
\begin{equation}
    \label{eq:S_tdlimit_longrange}
\lim_{L\to\infty}S_{\rm DW}(t)=e^{-\frac{t^2}{2}\left(\sum_{l=1}^p l u^2_l\right)}\,,
\end{equation}
which reduces to \eqref{eq:S_tdlimit} when $p=1$ and $u_1=-1/2$ , as expected. The spread complexity after the domain wall local quench with long-range evolution Hamiltonian can be obtained by adapting the results discussed above and reads
\begin{equation}
    \label{eq:spreadComplexity_DW_longrange}
C(t)=t^2\sum_{l=1}^p l u^2_l\,.
\end{equation}
As physical intuition would suggest, \eqref{eq:spreadComplexity_DW_longrange} implies that the more long-range hopping terms we involve in the evolution, the more the prefactor of its complexity grows.

\bibliographystyle{nb}
\bibliography{refs}

\end{document}